\RequirePackage{fix-cm}
\documentclass[twocolumn]{svjour3}          
\smartqed  
\usepackage{graphicx}
\usepackage{times}    
\mathchardef\bigtilde="0365

\begin{document}

\title{Estimates of critical quantities from an expansion in mass:\\
Ising model on the simple cubic lattice
}


\author{Hirofumi Yamada
}


\institute{H. Yamada \at
             Division of Mathematics and Science, Chiba Institute of Technology, 
\\Shibazono 2-1-1, Narashino, Chiba 275-0023, Japan\\
              \email{yamada.hirofumi@it-chiba.ac.jp}           
}

\date{Received: date / Accepted: date}

\maketitle

\begin{abstract}
In the Ising model on the simple cubic lattice, we describe the inverse temperature $\beta$ and other quantities relevant for the computation of critical quantities in terms of a dimensionless squared mass $M$.  The critical behaviors of those quantities are represented by the linear differential equations with constant coefficients which are related to critical exponents.  We estimate the critical temperature and exponents via an expansion in the inverse powers of the mass under the use of $\delta$-expansion.  The critical inverse temperature $\beta_{c}$ is estimated first in unbiased manner and then critical exponents are also estimated in biased and unbiased self-contained way including $\omega$, the correction-to-scaling exponent, $\nu$, $\eta$ and $\gamma$.
\end{abstract}


\section{Introduction}
Since the invention of lattice field theories, the border between condensed matter models and field theoretic models is lost and the two fields of physics are unified \cite{wil,kog}.  Traditionary, the both systems are described in terms of $\beta$ which indicates the inverse temperature or the inverse bare coupling constant.   In the field theory side, we notice that the lattice spacing $a$ or the equivalent full-mass can play the role of the basic parameter describing the models through the dependence of the bare coupling constant on $a$.  This feature naturally appears in the large $N$ limit of field theoretic models \cite{yam1}.  Also for finite-$N$ case of non-linear sigma models, it was shown that the scaling behavior of the inverse bare coupling has been captured from an expansion method in inverse powers of the full mass \cite{yam2}.   Then, it is a natural step to take the reverse point of view into the condensed matter models.  That is, from mathematical analogy of field theory models, the inverse temperature $\beta$ may be effectively described in the correlation length or the corresponding dimension-less mass parameter.  The behaviors of the inverse temperature and other thermodynamic quantities (the magnetic susceptibility etc) would be described by the mass.  The critical point is universally specified by the massless point and the scaling behavior described in the mass may emerge as the asymptotic behavior in the vicinity of the point.

In the present paper, we consider the 3-dimensional (3D) Ising model on the simple cubic lattice and try to estimate critical quantities, the critical temperature and critical exponents by the use of sole information in the high temperature expansion which is equivalent with the expansion in the inverse powers of the mass appropriately chosen.  To the best of our knowledge, most of the unbiased and self-contained computation has been carried out in field theoretic models and computer simulations.  For instance in Ising universality class at three dimension, there exist only a few literatures \cite{gut,butera1} where pure series expansions (high temperature expansions without universality hypothesis) of Ising models themselves were employed for the full exponents computation in a self-contained manner.  As the unbiased analysis for the first time, Guttmann studied the $19$th order high temperature series and obtained the estimates of $\beta_{c}$, $\gamma$ and $\nu$ on three 3D lattices \cite{gut} by the use of differential approximants method \cite{da}.  After a decade, using the same method, Butera and Comi investigated $21$st order high temperature series for the $N$ vector model and estimated those three quantities \cite{butera1}.  
 In general, accurate estimation of critical quantities becomes possible under the bias of the inverse critical temperature $\beta_{c}$ and/or the correction-to-scaling exponent $\theta$ obtained in other methods.   It is desirable, however, to equip necessary tools in one computation framework and we like to propose a self-contained approach to critical quantities in the expansion in a mass parameter. 
   
We adopt as the basic argument the second moment mass $M$ defined through the susceptibility $\chi=\sum_{j}<s_{0}s_{j}>$ and the second moment $\mu=\sum_{j}j^2<s_{0}s_{j}>$ where $s_{j}$ denotes the Ising spin at the site $j=(j_{1},j_{2},j_{3})$.   In more detail, consider the expansion of the two point function $\sum_{j}e^{-ipj}<s_{0}s_{j}>$ at low momentum $p$.  Using the expansion $e^{-ipj}=1-ipj-\frac{1}{2}(pj)^2+\cdots$, we obtain $\sum_{j}e^{-ipj}<s_{0}s_{j}>=\chi-\frac{\mu}{2D}p^2+O(p^4)=(2D\chi^2/\mu)/(2D\chi/\mu+p^2+O(p^4))$ ($D=3$).   Thus the squared mass $M$ is given by
\begin{equation}
M=\frac{2D\chi}{\mu}=\frac{6\chi}{\mu}.
\label{mass-def}
\end{equation}
In the high temperature expansion, $\chi$ and $\mu$ can be expanded in powers of $\beta$ and the inversion of $M(\beta)$ gives the series expansion of $\beta$ in $1/M$.  
  The key ingredient is the function $\beta$ expressed by $M$.  The analogous relation in field theories is the inverse bare coupling expressed by $a\Lambda$ where $\Lambda$ means a finite mass scale.  The dimensionless mass $M$ plays the role of the square of $a\Lambda$. 
  
From the form of the asymptotic expansion near the critical point, we consider that the critical behaviors of $\beta(M)$ and also other quantities are endowed with linear differential equations (LDE) with constant coefficients \cite{yam3}.   The LDE describing the scaling law in terms of $M$ is not satisfied by $\beta(M)$ at large $M$.  On the other hand our information available is the $1/M$ expansion of $\beta(M)$ with a certain convergence radius.  Thus, it is not allowed to use LDE to the $1/M$ series.  However, the so-called $\delta$-expansion is expected to change the status drastically \cite{yam1}.  Let us denote a function of interest be $f(M)$ and its truncated $1/M$ expansion, $f_{>}(M)$.  As well, $f_{<}(M)$ denotes the asymptotic expansion valid in the vicinity of the critical point $M=0$.  Then, it has been verified in some models that $\delta$-expansion induces a transform of $f_{>}(M)$ to $\bar f_{>} (t)$, and $\bar f_{>} (t)$ is indeed effective to investigate the critical behavior of $f(M)$, which can be confirmed by the plots of $\bar f_{>} (t)$ and its derivatives.   In other words, there is an overlapping between the effective regions of $\bar f_{>} (t)$ and $\bar f_{<} (t)$, and the critical quantities may be extracted by the LDE expected to be satisfied by $\bar f_{>} (t)$ at the matching region of $\bar f_{<}(t)$ and $\bar f_{>}(t)$.  For a detailed analysis on the 2D square Ising model, see ref. \cite{yam3}.  
After the confirmation of the scaling region realized in $\bar f_{>}(t)$ when necessary, we estimate critical quantities $\beta_{c}$,  $\nu$, $\omega$ ($=\theta/\nu$), $\eta$ and $\gamma$ in a fully self-contained way.  

This paper is organized as follows:  In section 2, to be self-contained, we briefly review the method of $\delta$  expansion on the lattice.   In section 3, we shall estimate $\beta_{c}$ from an expansion in the mass $M$ in an unbiased manner.  Rough estimation of $\nu$ is also obtained.  In section 4, we investigate in detail an improved estimate of critical exponents.   First, we estimate $\nu$ and $\eta$ by making use of the result $\omega\sim 0.84$ summarized in ref. \cite{peli}.   Then, unbiased and self-contained estimation of the exponents $\omega$, $\nu$, $\eta$ and $\gamma$ is attempted.   The last section is devoted to concluding remarks.

\section{$\delta$-expansion}
Let a given thermodynamic quantity as a function of $M$ be $f(M)$ and has the limit, $\lim_{M\to 0}f(M)=f_{c}$.  Then, our task is to estimate $f_{c}$ via the $1/M$ expansion of $f$ to the order $N$ denoted by $f_{>N}=\sum_{n=0}^{N}a_{n}M^{-n}$.   We change the variable from $M$ to $t$ by
\begin{equation}
M=\frac{1-\delta}{t},\quad (0\le \delta\le 1),
\label{mt}
\end{equation}
and view the function in $t^{-1}$ space where the critical region is enlarged by the factor $1-\delta$.  Then, we expand the function $f_{>N}((1-\delta)/t)$ in powers of $\delta$.  
Typical term $M^{-n}$ in $f_{>N}$ becomes as
\begin{equation}
M^{-n}=\Big(\frac{t}{1-\delta}\Big)^n=t^{n}\Big(1+n\delta+\frac{n(n+1)}{2}\delta^2+\cdots\Big).
\end{equation}
By truncating the $\delta$-expansion at $\delta^L$ and putting $\delta=1$, we obtain
\begin{equation}
M^{-n}\to \frac{(L+n)!}{n!L!}t^n.
\end{equation}
Thus, the $\delta$-expansion of $f_{>N}$ to order $L$ gives
\begin{equation}
f_{>N}\to \sum_{n=0}^{N}a_{n}\frac{(L+n)!}{n!L!}t^n.
\end{equation}
Empirically, it is found that the truncation order $L$ gives best result when $L$ depends on $N$ such that \cite{yam1,yam4}
\begin{equation}
L=N-n.
\label{presc}
\end{equation}
Throughout this paper we employ the prescription (\ref{presc}).  
Then
\begin{equation}
M^{-n}\to {N \choose n}t^{n},
\label{prescription}
\end{equation}
with the binomial factor
\begin{equation}
{N \choose n}=\frac{N!}{n!(N-n)!}=\frac{\Gamma(N+1)}{\Gamma(n+1)\Gamma(N-n+1)}.
\end{equation}
We notice that ${N \choose 0}=1$ and $M^{0}\to  t^{-0}$.  The $\delta$-expansion keeps a constant term invariant.  

Thus, we define the $\delta$-expansion of $f_{>N}(M)$ by
\begin{equation}
f_{>N}(M)\to \sum_{n=0}^{N}a_{n}{N \choose n}t^{n}=:D_{N}[f_{>N}(M)]=:\bar f_{>N}(t).
\label{borel}
\end{equation}
To summarize, the $\delta$-expansion creates a new function $\bar f_{>N}(t)$ associated with $f_{>N}(M)$ by transforming the coefficient from $a_{n}$ to $a_{n}{N \choose n}$ which depends on the order $N$.  The symbol $D_{N}$ denotes the order dependent transformation from $f_{>N}(M)$ to $\bar f_{>N}(t)$.   Since ${N \choose n}\to \frac{N^n}{n!}$ ($n$:fixed, $N\to \infty$), one might think that the series becomes ill-defined in the $N\to \infty$ limit.  However, it is shown in some simple models that there exists a region of the form $(0,t_{0})=I$ where $\bar f_{>N}(t)$ develops a plateau there and converges almost everywhere to $f_{c}$ as $N\to \infty$ \cite{yam3}.  This is conceivable, since putting $\delta=1$ means $M=0$ for every $t$ (see (\ref{mt})).  We thus assume that, when  $\lim_{M\to 0}f(M)=f_{c}$ exists,
\begin{equation}
\lim_{N\to \infty}\bar f_{>N}(t)=f_{c}
\label{clue}
\end{equation}
over a certain region $I$.  As explicitly studied in ref. \cite{yam3}, in one-dimensional Ising model, the above result is analytically confirmed.  In the square Ising model, numerical analysis to large orders of $1/M$ expansion has also served us the convincing result supporting (\ref{clue}). In the case of simple cubical lattice, we can confirm (\ref{clue}) by the direct numerical experiment for relevant functions as shown in various plots (see, for example,  the plots of $D_{N}[\beta_{>N}]=\bar \beta_{>N}$ in Fig. 1).    The point is, in the $t$-space, the whole region of $I$ becomes the scaling region as long as the order $N$ is high enough.  We emphasize that the effective region does not include the origin $t=0$ inside.

In the presence of phase transition, we must deal with the case where the expansion of $f$ around $M=0$ is not regular.  Then we consider how in such a case the $\delta$-expansion affects the small $M$ behavior of $f(M)$,  supposing that $f$ behaves at small enough $M$ as $f\sim f_{c}+f_{1}M^{\alpha_{1}}+f_{2}M^{\alpha_{2}}+\cdots=:f_{<}$ where $0<\alpha_{1}<\alpha_{2}<\cdots$.  When $M=t^{-1}(1-\delta)$ is substituted into $M^{\alpha}$ and $((1-\delta)/t)^{\alpha}$ is expanded in $\delta$, giving
$
M^{\alpha}=t^{-\alpha}(1+\alpha \delta+\cdots),
$
a reasonable truncation protocol for the best matching with $\bar f_{<N}(t)$ is not found on logical grounds yet.   Here, we proceed by simply extending the factor (\ref{prescription}) by the extension of $\Gamma$ functions, 
\begin{equation}
M^{\alpha}\to {N \choose -\alpha}t^{-\alpha},
\end{equation}
where
\begin{equation}
{N \choose -\alpha}=\frac{\Gamma(N+1)}{\Gamma(-\alpha+1)\Gamma(N+\alpha+1)}.
\label{ansatz}
\end{equation}
We thus define the $\delta$-expansion of $f_{<}(M)$ is given by
\begin{equation}
D_{N}[f_{<}(M)]=f_{c}+f_{1}{N \choose -\alpha_{1}} t^{-\alpha_{1}}+\cdots.
\end{equation}   The generalized binomial factor ${N \choose -\alpha}$ vanishes when $\alpha=1,2,3,\cdots$ and
\begin{equation}
D_{N}[M^{\alpha}]=0,\quad (\alpha=1,2,3,\cdots).
\end{equation}
This means that the regular contributions in the scaling region are essentially eliminated by the $\delta$ expansion.  
For non-integer positive power of $M$, the factor decreases with the order and disappears in the $N\to \infty$ limit,
\begin{equation}
{N \choose -\alpha}\to \frac{N^{-\alpha}}{\Gamma(-\alpha+1)}\to 0,\quad  (N\to \infty).
\end{equation}
As the exponent $\alpha$ is large, the factor decreases faster.   All these facts appeal the advantage of $\bar f_{N}(t)$ compared to $f_{N}(M)$.   

In the present work, our task is summarized to estimate $f_{c}$ and $\alpha_{k}$ $(k=1,2,\cdots)$ from the known series (\ref{borel}).  In the process, we use derivatives of $f_{N}$ ($=f_{>N}\,\,{\rm or}\, \,f_{<N})$.  To this end we remark that
\begin{equation}
t\frac{d}{dt}D_{N}[f_{N}(M)]=D\Big[(1/M)\frac{d}{d(1/M)}f_{N}(M)\Big],
\end{equation}
which states that $D_{N}$-operation and differentiation is commutable.  It is convenient to use the following abbreviate notation,
\begin{equation}
\Big(t\frac{d}{dt}\Big)^{k}\bar f_{N}(t)=\bar f_{N}^{(k)}(t),\quad (k=0,1,2,3,\cdots).
\end{equation}
From (\ref{borel}) and the property of the binomial factor, we obtain
\begin{equation}
\bar f_{>N}^{(1)}=N(\bar f_{>N}-\bar f_{>N-1}).
\label{dif}
\end{equation}
This relates the derivative to the backward difference of the functions with respect to $N$.  
It means that the stationary point of $\bar f_{>N}$ sits on the curve of $\bar f_{>N-1}$, one lower order function.  The second order derivative obeys similar constraint,
\begin{equation}
\bar f_{>N}^{(2)}=N(N-1)(\bar f_{>N}-2\bar f_{>N-1}+\bar f_{>N-2})+\bar f_{>N}^{(1)}.
\label{2dif}
\end{equation}
Hence, when $\bar f_{>N}^{(1)}=\bar f_{>N}^{(2)}=0$ at $t=t^*$, it follows that $\bar f_{>N}=\bar f_{>N-1}=\bar f_{>N-2}$ at $t=t^*$.  In general, (\ref{dif}), (\ref{2dif}) and their higher order followers state on the convergence issue of the sequence of estimates.  That is, if a point satisfying approximately the stationarity to the higher orders, the estimates of $f_{c}$ by that stationary point ensures the approximate coincidence of estimations at successive orders, $f_{c}(N)$, $f_{c}(N-1)$, $\cdots$, showing good convergence behavior. 

\section{Estimation of $\beta_{c}$ and $\nu$}
The Ising model on the simple cubic lattice is defined by the action
\begin{equation}
S=-\beta \sum_{< i,j>} s_{i}s_{j},\quad s_{i}^2=1,
\end{equation}
where the spin sum is over all nearest neighbour pair $<i,j>$ on the periodic lattice.  
Our purpose here is to estimate the inverse critical temperature $\beta_{c}$ defined in our approach by
 \begin{equation}
 \beta_{c}=\lim_{M\to 0}\beta(M).
 \end{equation}
 
Past studies up to 2002 year on the estimation of $\beta_{c}$ and exponents are 
reviewed by Pelissetto and Vicari \cite{peli} in comprehensive manner.  Of course there are important new contributions after ref. \cite{peli} on the subject of accurate estimations of the critical quantities \cite{ari,den,ber,pog,has,lit,gor,gli,she,harada}.  
The best estimations of $\beta_{c}$ to our knowledge before 2002 are $\beta_{c}= 0.22165459(10)$ \cite{blo}, $\beta_{c}= 0.2216595(15)$ \cite{ito}, and the recent works added $\beta_{c}= 0.2216545(1)$ and $0.2216542(4)$ \cite{ari}, $\beta_{c}=0.22165455(3)$ \cite{den}, $\beta_{c}= 0.22165463(8)$ \cite{has} and $\beta_{c}= 0.221652(2)$ \cite{harada}.  Newer estimations agree with each other to the first $6$ decimal places and, in this work, we refer the range of $\beta_{c}$ as the modest one,
\begin{equation}
\beta_{c}=0.22165\sim 0.22166,
\label{betac}
\end{equation}
within which all recent estimates are included.  The world average of $\nu$ up to the review \cite{peli} is given by $\nu=0.6301(4)$.  Newer estimates are $\nu=0.63002(10)$ \cite{has}, $\nu=0.630(5)$ \cite{lit}, $\nu=0.6306(5)$ \cite{pog}, $\nu=0.63048(32)$ \cite{gor},  $\nu=0.629(1)$ \cite{gli}, $\nu=0.62999(5)$ \cite{she} and $\nu=0.6301(8)$ \cite{harada}.  Now, though most accurate estimates come from \cite{has} and \cite{she}, we refer
\begin{equation}
\nu=0.6301(4)
\label{nu_s}
\end{equation}
summarized in \cite{peli}, since new results support (\ref{nu_s}).

\subsection{Preliminary studies}
Our approach is based upon the high temperature expansion.  The magnetic susceptibility and the second moment have been computed up to $\beta^{25}$ by Butera and Comi \cite{butera}.  From the definition of the second moment mass squared (\ref{mass-def}) and the result reported in \cite{butera}, $\chi_{>}=1+6\beta+30\beta^2+O(\beta^3)$ and $\mu=6\beta+72\beta^2+O(\beta^3)$, we find $1/M=\beta+6\beta^2+\frac{92}{3}\beta^3+O(\beta^4)$.  By the simple inversion of the series, we then obtain
\begin{eqnarray}
\beta_{>}&=&x-6x^2+\frac{124}{3}x^3-312x^4+\frac{12596}{5}x^5-21432 x^6\nonumber\\
& &+\frac{1330848}{7}x^7-1745344 x^8+\frac{148384348}{9}x^9\nonumber\\
& &-\frac{797787336}{5}x^{10}+\frac{17341288504}{11}x^{11}\nonumber\\
& &-15857888272 x^{12}+\frac{2106367479672}{13}x^{13}\nonumber\\
& &-\frac{11748802870160}{7}x^{14}+\frac{263968267347944}{15}x^{15}\nonumber\\ & &-186504592354608 x^{16}+\frac{33924951987330804}{17}x^{17}\nonumber\\
 & &-21535692193295224 x^{18}\nonumber\\
& &+\frac{4449606807205690200}{19}x^{19}\nonumber\\
& &-\frac{12821205881021198992}{5}x^{20}\nonumber\\
& &+\frac{197756701920466780928}{7}x^{21}\nonumber\\
& &-\frac{3442869826889278353376}{11}x^{22}\nonumber\\
& &+\frac{80156432259652309452520}{23}x^{23}\nonumber\\
& &-\frac{116948936021276297965072}{3}x^{24}\nonumber\\
& &+\frac{10946582972904015563857296}{25}x^{25}+O(x^{26}),
\label{beta_M}
\end{eqnarray}
where
\begin{equation}
x:=\frac{1}{M}.
\end{equation}
The result of the $\delta$-expansion to the order $N$ is readily obtained by multiplying the binomial factor to the $n$th order coefficient $b_{n}$, giving
\begin{equation}
\bar\beta_{>N}:=D_{N}[\beta_{>N}]=\sum_{n=1}^{N}b_{n}{N \choose n}t^n.
\label{delbeta}
\end{equation}

Now, we turn to argue the behavior of $\beta$ near the critical point to prepare our estimation work.  In the study of phase transition via series expansion technique, we rely upon heuristic assumption of the power law near transition point.  
In reference to \cite{peli}, we start the arguments by the consideration of the Wegner expansion \cite{weg} of the correlation length $\xi$ at temperatures near $1/\beta_{c}$ in terms of the reduced temperature $\tau=1-\beta/\beta_{c}$.  Let the spectrum of powers of corrections be generated by the basic ones, $1$, $\theta$, $\theta_{1}$, $\theta_{2}$, $\cdots$ and their integer multiples.  Then, 
\begin{eqnarray}
\xi&\sim& f\tau^{-\nu}\bigg[1+\Big\{a\tau^{\theta}(1+a_{11}\tau+a_{12}\tau^2+\cdots)\nonumber\\
& &+a_{2}\tau^{2\theta}(1+a_{21}\tau+a_{22}\tau^2+\cdots)+\cdots\Big\}\nonumber\\
& &+\Big\{b\tau^{\theta_{1}}(1+b_{11}\tau+b_{12}\tau^2+\cdots)\nonumber\\
 & &+b_{2}\tau^{2\theta_{1}}(1+b_{21}\tau+b_{22}\tau^2+\cdots)+\cdots\Big\}\label{xi}\\
 & &+\cdots\cdots+\Big\{u\tau(1+u_{1}\tau+u_{2}\tau^2+\cdots)\Big\}\bigg]+\xi_{R},\nonumber
\end{eqnarray}
where $\xi_{R}=const\times \tau(1+r_{1}\tau+r_{2}\tau^2+\cdots)$ stands for the analytic back ground  and $0<\theta<\theta_{1}<\theta_{2}<\cdots$.   The inversion yields a series in $\xi/f$ with the spectrum of exponents of the form scaled by $\nu$, $-1/\nu-(m+n\theta+n_{1}\theta_{1}+n_{2}\theta_{2}+\cdots)/\nu$ where $m, n, n_{k}\, (k=1,2,3,\cdots)$ are all non-negative integers.  Leading term is $(\xi/f)^{-1/\nu}$ provided $\theta<1$ and the next term is $(\xi/f)^{-(1+\theta)/\nu}$.  Here, we pose the assumption that $\theta_{1}>1$ and then the third term proves to be $(\xi/f)^{-(1+1)/\nu}=(\xi/f)^{-2/\nu}$ (provided $\theta>1/2$).  Then, inversion of the above series to a few orders reads
\begin{equation}
\tau=(\xi/f)^{-1/\nu}+\frac{a}{\nu}(\xi/f)^{-(1+\theta)/\nu}+\frac{b}{\nu}(\xi/f)^{-2/\nu}+\cdots.
\end{equation}

The correlation length $\xi$ is defined through the large separation limit of the two point function $<s_{0}s_{j}>$ ($|j|\to \infty$) and therefore has relation with the mass $M$  defined in (\ref{mass-def}).  Actually, it is known that for the case when the site $j$ is on one axis, $M=2(\cosh \xi^{-1}-1)$ \cite{tarko}.  Thus, 
 $\xi^{-1}\sim M^{1/2}(1+O(M))$ near the critical point and $\xi^{-1}\sim \log M$ in the high temperature limit ($M\to \infty$).  Though the high temperature behavior may be $\xi^{-1}\sim const\times \log(const \times M)$ for the general position of the site $j$ (One can understand this by directly performing the high temperature expansion of $<s_{0}s_{j}>$ at the leading order \cite{comment}), the relation near the critical point should be universally given by $\xi^{-1}\sim M^{1/2}(1+O(M))$, since the result must become independent of the direction of the site $j$.   Thus, near the massless limit, $\xi^{-1/\nu}\sim M^{1/2\nu}(1+O(M))$ and the term of order $M^{1/2\nu+1}$ appears in $\tau$.  This term belongs to rather higher orders in our study and thus we arrive at
\begin{equation}
\tau=f^{\frac{1}{\nu}}M^{\frac{1}{2\nu}}(1+\frac{a}{\nu}f^{\frac{\theta}{\nu}}M^{\frac{\theta}{2\nu}}+\frac{b}{\nu}f^{\frac{1}{\nu}}M^{\frac{1}{2\nu}}+\cdots),
\label{tauscale}
\end{equation}
and 
\begin{equation}
\beta=\beta_{c}(1-\tau).
\label{beta_scaling}
\end{equation}

Having finished a survey on the series, let us start the argument by supposing the scaling behavior of inverse temperature in the power-like form.  By the combination of (\ref{tauscale}) and (\ref{beta_scaling}), we write
\begin{equation}
\beta_{<}=\beta_{c}-A_{1} x^{-p_{1}}-A_{2} x^{-p_{2}}-A_{3} x^{-p_{3}}-O(x^{-p_{4}}),\label{betascaling}
\end{equation}
where $0<p_{1}<p_{2}<p_{3}<\cdots$ and
\begin{equation}
\bar\beta_{<}=\beta_{c} -\bar A_{1} t^{-p_{1}}-\bar A_{2} t^{-p_{2}}-\bar A_{3} t^{-p_{3}}-O(t^{-p_{4}}),
\label{delbetascaling}
\end{equation}
where
\begin{equation}
\bar A_{n}=A_{n}{N \choose -p_{n}}.
\end{equation}
We note that
\begin{equation}
p_{1}=\frac{1}{2\nu},\quad p_{2}=\frac{1+\theta}{2\nu},\quad p_{3}=\frac{1+1}{2\nu}.
\end{equation}  
Since ${N \choose -p_{n}}\to 0$ as $N\to \infty$, we may assume that over a certain region of $t$, 
\begin{equation}
\lim_{N\to \infty}\bar\beta_{<N}=\beta_{c}.
\label{clue2}
\end{equation}

\begin{figure}
\centering
\includegraphics[scale=0.8]{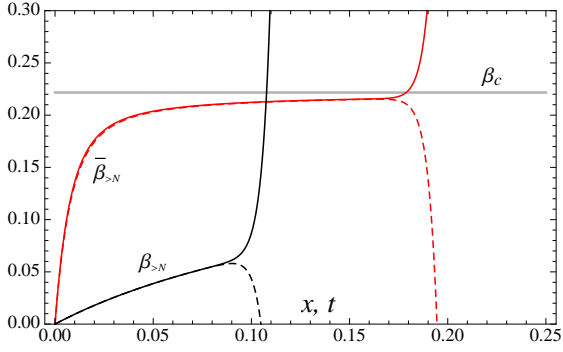}
\caption{(Color online) Plots of $\beta_{>N}(x)$ (black) and $\bar\beta_{>N}(t)$ (red) at $N=24$ (dashed), $25$ (solid).  It is shown that $\bar\beta_{>N}$ indicates the critical temperature $\beta_{c}=0.22165\cdots$ in the plateau region.  In $\bar\beta_{>N}(t)$, the limit of effective region may be signaled by the point of least variation and at that point the value of $\beta_{c}$ is best estimated.}
\end{figure}

The effects of $\delta$-expansion are clearly shown in the plots of relevant functions.  For an example, the plot of $\beta_{>25}(x)$ and $\bar\beta_{>25}(t)$ in Fig. 1 shows that $\bar\beta_{>25}(t)$ exhibits the trend of approaching to the limit $\beta_{c}$, while $\beta_{>25}(x)$ has no sign of the scaling, as it would.   This numerical examination provides a clear evidence that the $\delta$-expansion gives effective function of $t$ capturing the scaling behavior at non-large $t$.  
\begin{figure}[t]
\centering
\includegraphics[scale=0.8]{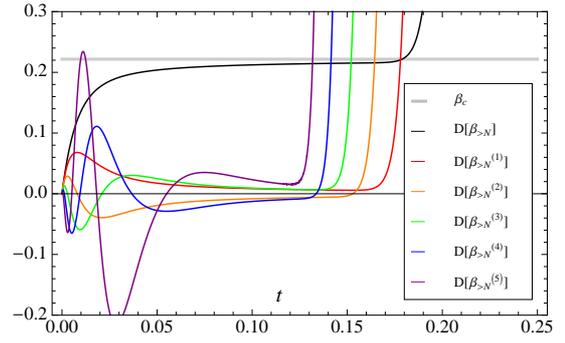}
\caption{(Color online) Plots of $\bar\beta_{>25}$ and $\bar\beta_{>25}^{(k)}$ $(k=1\sim 5)$.   The high temperature region is pushed away to the neighborhood of the origin $t=0$ and the scaling region appears around $t\sim 0.11$ for all plotted derivatives.}
\end{figure}
We also show in Fig. 2 the plots of $\bar\beta_{>25}$ and $\bar\beta_{>25}^{(k)}$ for $k=1\sim 5$.   Also in the derivatives, it seems that the scaling behaviors have emerged;   After oscillations of a few or several times, the derivatives set in their scaling behaviors of monotonically approaching to zero.   It however needs a detailed study to judge to what level or grade the observed behaviors can be classified.   On this respects, logarithmic plots of the derivative $\bar\beta^{(k)}(t)$ $(k=1,2,3,4)$ help the assessment.  The log-plots of derivatives using $\bar\beta_{>}^{(k)}(t)$ effective at small $t$ including the square model are shown in Fig. 3.  The shown behaviors should be compared with the following,
\begin{equation}
\log((-1)^{k+1}\bar\beta_{<}^{(k)})=\log(p_{1}^k \bar A_{1})-p_{1}\log t+O(t^{-p_{2}+p_{1}})
\label{derivative}
\end{equation}
\begin{figure}[t]
\centering
\includegraphics[scale=0.8]{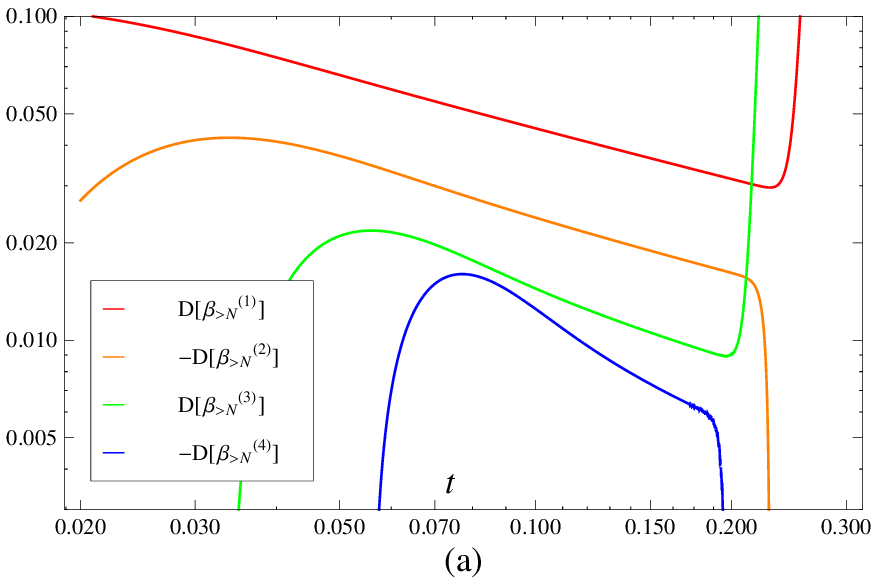}
\includegraphics[scale=0.8]{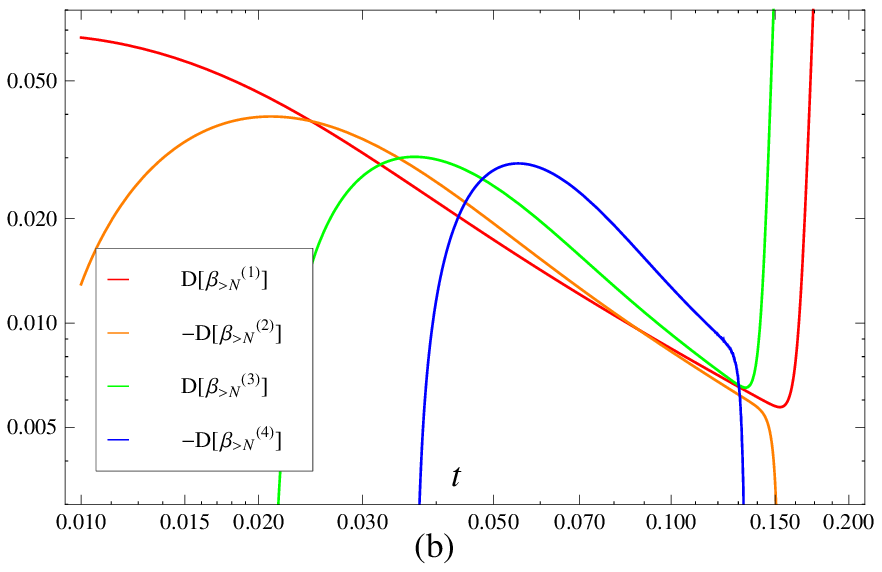}
\caption{(Color online) Logarithmic Plots of $\bar\beta_{>25}^{(k)}$ $(k=1\sim 4)$ in (a) the square Ising model and (b) the cubic Ising model at $N=25$.}
\end{figure}
We note that the linear-like behavior in the logarithmic plot is necessary for a benchmark of the scaling.  If the behavior of $\log((-1)^{k+1} \bar\beta_{>N}^{(k)})$ is linear and the correction $O(t^{-p_{2}+p_{1}})$ is negligible, then the leading terms are dominant and the scaling behavior is pure and high grade.  In this case, the gradient yields a good approximation for $-p_{1}=-1/(2\nu)$.  If the behavior is linear-like but the correction is still affective, the scaling behavior is a mixture of the leading and correction terms.
  
The square model at $25$th order exhibits clear and pure scalings for the first and second derivatives, as shown in Fig. 3a (We have used an exact mass $M=2(\cosh \xi^{-1}-1)$ \cite{tarko} where the correlation length $\xi$ is given by $\beta$ as $\xi^{-1}=-\log\tanh\beta-2\beta$ \cite{mont}).  The points appropriate for the linearization of the first derivative are at $t=0.172647\cdots$ and $0.210147\cdots$.  The slope values at respective points are $-0.515381\cdots$ and $-0.508829\cdots$, respectively.   As for the second derivative, the linearization is suitable at $t=0.190590\cdots$ and the gradient takes the  value $-0.544483\cdots$.  As expected, the value of the gradient in the first derivative at larger $t$ is most accurate to the exact value $-1/(2\nu)=-0.5$.  The third order derivative shows rather mixed scaling and the fourth order, weak and obscure.  Thus, the times of differentiation increases, the scaling behavior becomes less-qualified.   

At 3-dimension, the scaling behavior is not so high graded compared to the 2-dimensional case.   For instance, though the curve of the third order derivative must be in the lower region of the second order one, it is located at the upper region of the second order one.  Thus, the trend of the scalings are indeed implied by the linear-like behavior, but the level of the scaling is not so high except for the first order.  
The vertical intervals between the derivatives are narrower at 3D compared with the 2D case.  This is due to the fact that the interval roughly given by $-\log p_{1}$ is smaller for 3D model implying the closeness of $p_{1}=1/(2\nu)$ to unity.

We emphasize that, the $\delta$-expansion is still effective to uncover the scaling region within small $t$ regime.  In particular for $\bar\beta_{>N}$, it is obtained as the integration of its first derivative and the scaling behavior is expected to be better.   
The problem is that the leading correction of order $O(t^{-p_{1}})$ is not negligible in $\bar\beta_{>N}$ yet and the estimate by the stationary point in the plateau does not lead high accuracy for $N\le 25$.  Now then, our idea upon the estimation of $\beta_{c}$ is to cancel dominant corrections to $\beta_{c}$ in (\ref{delbetascaling}) by incorporating the derivative and consider the combination,
\begin{equation}
\Big[1+p_{1}^{-1}\frac{d}{d\log x}\Big]\beta_{<}.
\end{equation}
Then it holds approximately at large $x$ (small $M$) that
\begin{equation}
\Big[1+p_{1}^{-1}\frac{d}{d\log x}\Big]\beta_{<}\sim \beta_{c}.
\label{dif1}
\end{equation}
The discrepancy between both sides is of order $O(x^{-p_{2}})$ and reduced to the second order correction.

From (\ref{clue}) again, for large $N$, $\bar\beta_{<N}$ would approximately obey (\ref{dif1}) over a certain region that 
\begin{equation}
\Big[1+p_{1}^{-1}\frac{d}{d\log t}\Big]\bar \beta_{<N}=\beta_{c}+O(t^{-p_{2}}),
\label{delpsi_conv}
\end{equation} 
where $\bar \beta_{<N}(t):=D_{N}[\beta_{<}(M)]$.   We notice that, as the extension,  one can readily build up $K$th order LDE such that
\begin{equation}
\prod_{k=1}^{K}\Big[1+p_{k}^{-1}\frac{d}{d\log t}\Big]\bar \beta_{<N}=\beta_{c}+O(t^{-p_{K+1}}),
\label{delpsi_conv_multi}
\end{equation}
for $K=1,2,3,\cdots$.  
For the sake of notational simplicity, we denote
\begin{equation}
\bar\Psi_{<}:=\prod_{k=1}^{K}\Big[1+p_{k}^{-1}\frac{d}{d\log t}\Big]\bar \beta_{<N}.
\end{equation}
Making use of (\ref{delpsi_conv}) or (\ref{delpsi_conv_multi}), we carry out estimation of $\beta_{c}$ and $\nu$ in the following subsections.

Closing comment on the naturalness of the scheme based upon LDE:  We note that (\ref{betascaling}) can be written as $\beta_{<}=\beta_{c}-A_{1} e^{-p_{1}\log x}-A_{2} e^{-p_{2}\log x}-\cdots$.  This form urges us to consider that the series is an expansion in eigen functions of the linear operator $\prod_{k}[p_{k}+d/d\log x]$.  When there exists a quantity in which some terms are multiplied by the logarithmic factor of $x$ giving $\log x\times  x^{-\lambda_{k}}$, it means that the eigen value $\lambda_{k}$ is doubly degenerated.  The magnetic susceptibility and the specific heat at 2D just correspond to such a case.   Also in field theoretic models, ordinary quantum corrections come as logarithmic corrections.  The  approach explored in this work would be capable of handling these cases.

\subsection{Estimate of $\beta_{c}$ and $\nu$ with the 1st order LDE}
The starting point is (\ref{delpsi_conv}) which states that $\bar\Psi_{<}=[1+p_{1}^{-1}\\(d/d\log t)]\bar \beta_{<N}$ approaches as $N\to \infty$ to $\beta_{c}$.  First, at finite $N$, we employ the principle of minimum sensitivity (PMS) \cite{steve}.    The use of the principle is quite natural since the convergence region of the left-hand-side of (\ref{delpsi_conv}) or (\ref{delpsi_conv_multi}) should appear as a plateau.  Thus, the value of $\bar\Psi_{<}\sim \bar\Psi_{>}=\bar\beta_{>N}+p_{1}^{-1}\bar\beta^{(1)}_{>N}$ at a possible stationary point gives an estimation of $\beta_{c}$.  To obtain numerical results, the value of $p_{1}=1/(2\nu)$ is necessary.  Hence, to fix all unknown quantities $\beta_{c}$ and $p_{1}$, we extend PMS condition such that the stationarity of $\bar\Psi_{>}$ becomes maximal at a proper value of $p_{1}$ by further imposing the second derivative $\bar\Psi_{>}^{(2)}=\beta_{>N}^{(2)}+p_{1}^{-1}\bar\beta^{(3)}_{>N}$ to vanish or be minimum in the magnitude (This extension was first proposed in ref. \cite{knp}).  Now, the point is to remind that $\bar \beta_{>N}$ exhibits the scaling behavior in some region.  From Fig. 1, we convince ourselves that the upper limit of the region showing the scaling behavior of $\bar\beta_{>25}^{(k)}$ $(k=0,1,2)$ is about $t\sim 0.15$.  Making use of this, we replace $\bar\beta_{<N}$ by $\bar\beta_{N>}$ in the estimation task under extended PMS.

Explicitly the extended PMS condition in terms of $\bar\beta_{>N}^{(k)}$ reads
\begin{eqnarray}
\bar\Psi_{>}^{(1)}=\bar\beta_{>N}^{(1)}+p_{1}^{-1} \bar\beta_{>N}^{(2)}&=&0,
\label{rho}\\
\bar\Psi_{>}^{(2)}=\bar\beta_{>N}^{(2)}+p_{1}^{-1} \bar\beta_{>N}^{(3)}&\sim &0.
\label{rho2}
\end{eqnarray}
The symbol $\sim $ in the second condition means the local minimality in the magnitude.  It is straightforward to obtain $p_{1}^{-1}$ from (\ref{rho}) as the function of $t$ which indicates the estimation point.  To avoid possible confusion we denote the solution as $\rho(t)$.  Thus, $\rho(t) =-\bar\beta_{>N}^{(1)}/\bar\beta_{>N}^{(2)}$.  Then, substituting it into (\ref{rho2}), we obtain $\bar\Psi_{>}^{(2)}=\bar\Psi_{>}^{(2)}|_{\rho(t)}=((\bar\beta_{>N}^{(2)})^2-\bar\beta_{>N}^{(1)}\bar\beta_{>N}^{(3)})/\bar\beta_{>N}^{(2)}$ and from $\bar\Psi_{>}^{(2)}|_{\rho(t)}\sim 0$ the point $t=t^{*}$ at which the stationarity is maximally realized is found.   Next, $\rho(t^{*})=1/p_{1}^{*}$ gives the estimate of $\nu$ by $\nu=1/(2p_{1}^{*})$.   In this manner, we have estimation of $\beta_{c}$ by
\begin{equation}
\beta_{c}=\bar\Psi_{>}\Big |_{t=t^{*}}=\bar\beta_{>N}(t^{*})+(1/p_{1}^{*}) \bar\beta_{>N}^{(1)}(t^{*}).
\end{equation}

We carry out this procedure from $4$th order to $25$th order.  The reliable and steady trend of our approach emerges from $13$th order.   The trend is the characteristic scaling behavior that the $\rho(t)$-substituted second derivative $\bar\beta_{>N}^{(2)}+\rho(t) \bar\beta_{>N}^{(3)}$ should exhibit, which we describe below:
\begin{figure}[ht]
\centering
\includegraphics[scale=0.75]{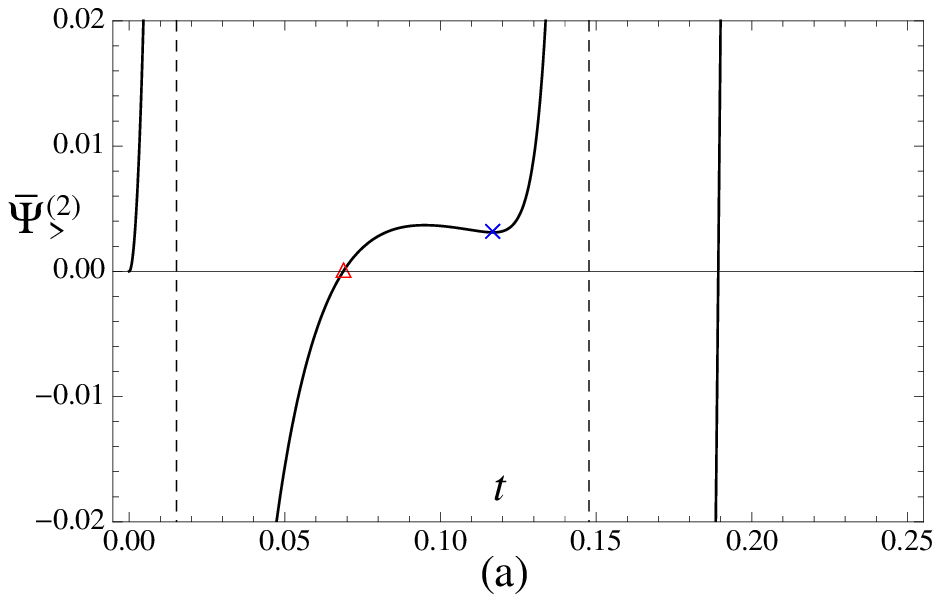}
\includegraphics[scale=0.75]{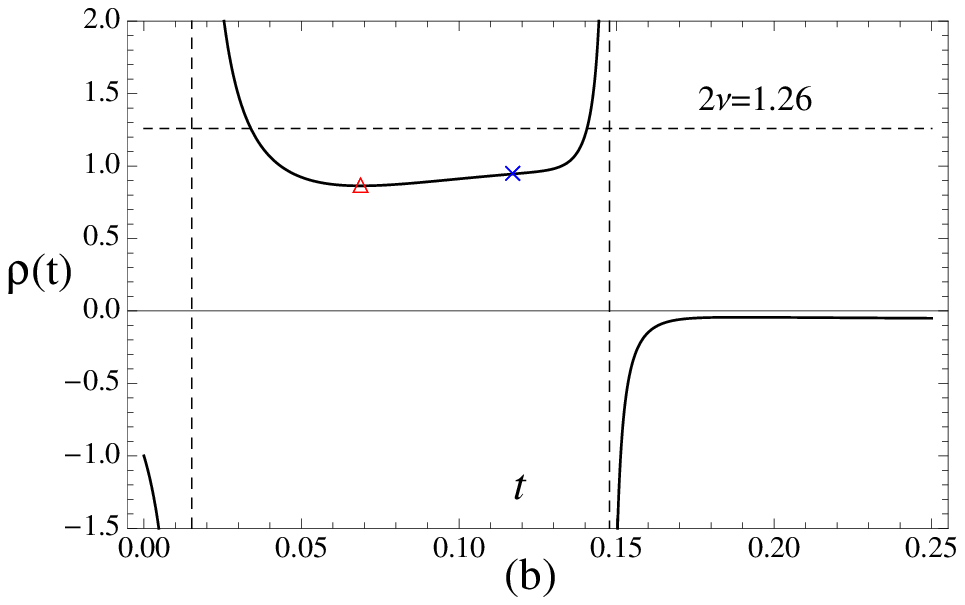}
\includegraphics[scale=0.75]{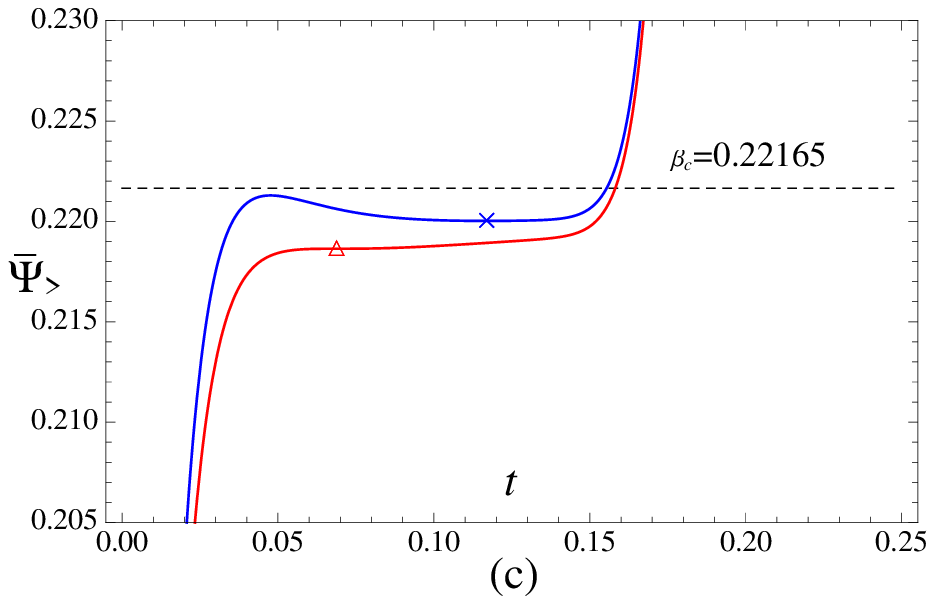}
\caption{(Color online) Plots of (a) $\bar\Psi_{>}^{(2)}(\rho(t), t)=((\bar\beta_{>N}^{(2)})^2-\bar\beta_{>N}^{(1)}\bar\beta_{>N}^{(3)})/\bar\beta_{>N}^{(2)}$, (b) $\rho(t)$ and (c) $\bar\Psi_{>}(\rho^{*}, t)$ at $N=13$.  Vertical dashed lines in (a) and (b) indicate the singular points coming from the solution of $\bar\beta_{N>}^{(2)}=0$.  They are at $t=0.01526$ and $0.14768$.  Note that between the two points, $\rho$ is positive definite.  The range of optimal stationary points are thus limited within $0.01526 <t^{*}< 0.14768$.  Two candidates of physical interest due to the extended PMS are indicated by the red triangle and blue cross.  Among them, important one is the solution represented by blue cross, which we call proper solution, because it lies in the place where the critical region begins to appear.  The lower plot shows two $\bar\Psi_{>}(\rho^{*}, t)$ at $N=13$.  The red one is for $\rho^{*}=0.863705$ and the blue one is for $\rho^{*}=0.945108$.}
\end{figure}
The solution $\rho(t)$ should behave at $t$ in the critical region as
\begin{eqnarray}
\rho(t)&=&\frac{\sum_{k=1}^{\infty} p_{k} \bar A_{k}t^{-p_{k}}}{\sum_{k=1}^{\infty}(p_{k})^2 \bar A_{k}t^{-p_{k}}}\nonumber\\
&=&\frac{1}{p_{1}}-\frac{p_{2}\bar A_{2}}{p_{1}^3 \bar A_{1}}(p_{2}-p_{1}) t^{-(p_{2}-p_{1})}+\cdots.
\end{eqnarray}
Then $\bar\Psi_{>}^{(2)}=\bar\beta_{>N}^{(2)}+\rho(t) \bar\beta_{>N}^{(3)}$ loses $t^{-p_{1}}$ term and has to behave as 
\begin{equation}
\bar\Psi_{>}^{(2)}=\frac{p_{2}}{p_{1}}(p_{1}-p_{2})^2 \bar A_{2}t^{-p_{2}}+\cdots.
\label{scaling_324}
\end{equation}
Note that $\rho(t)\to 1/p_{1}$ since ${N \choose -p_{2}}/{N \choose -p_{1}}\sim \Gamma(1-p_{2}) /\Gamma(1-p_{1}) \times N^{-p_{2}+p_{1}}$ tends to zero ($p_{1}<p_{2}$) as $N\to \infty$.  In the same manner, the correction in (\ref{scaling_324}) vanishes in the $N\to \infty$ limit.   At finite order, the correction $\sim const\times t^{-p_{2}}$ however remains.  For instance, as implied in Fig. 4(a), the weak peak of $\bar\Psi_{>}^{(2)}=\bar\beta_{>N}^{(2)}+\rho(t) \bar\beta_{>N}^{(3)}$ about $t\sim 0.09$ at $N=13$ is interpreted to show the transition from high temperature to the scaling region.   The dotted vertical lines in Fig. 4(a) and (b) represent the divergences of $\bar\Psi_{>}^{(2)}$ and $\rho(t)$ at $N=13$.   

As shown in Fig. 4 (a), (b), we obtain two candidates of the $\beta_{c}$ estimation.  One comes from the zero of $\bar\Psi_{>}^{(2)}=\bar\beta_{>N}^{(2)}+\rho(t) \bar\beta_{>N}^{(3)}$ and the other from its local minimum ($\neq 0$).  The first candidate is outside ($t\sim 0.07$) and the other is at the beginning ($t\sim 0.12$) of the scaling region.  The later one represented by the blue cross is also the point marking the upper limit of the reliable region of the truncated $\bar\Psi_{>}^{(2)}(t)$, $\rho(t)$ and $\bar\Psi_{>}(t)$.  Hence, we regard the candidate marked by the blue cross as important and optimal.  Actual values computed are $\beta_{c}=0.218638\, ({\rm at}\,\,t^{*}=0.068789)$ and $\beta_{c}=0.220024\, ({\rm at}\,\,t^{*}=0.116908)$.  To summarize, we can thus select which solution is the best choice among appeared.  We note that at the point where $\bar\Psi_{>}^{(2)}=0$, $\rho(t)$ becomes extremal.  At general order $N$, this is found by the direct differentiation of $\rho(t)$ by $\log t$.
\begin{figure}[ht]
\centering
\includegraphics[scale=0.8]{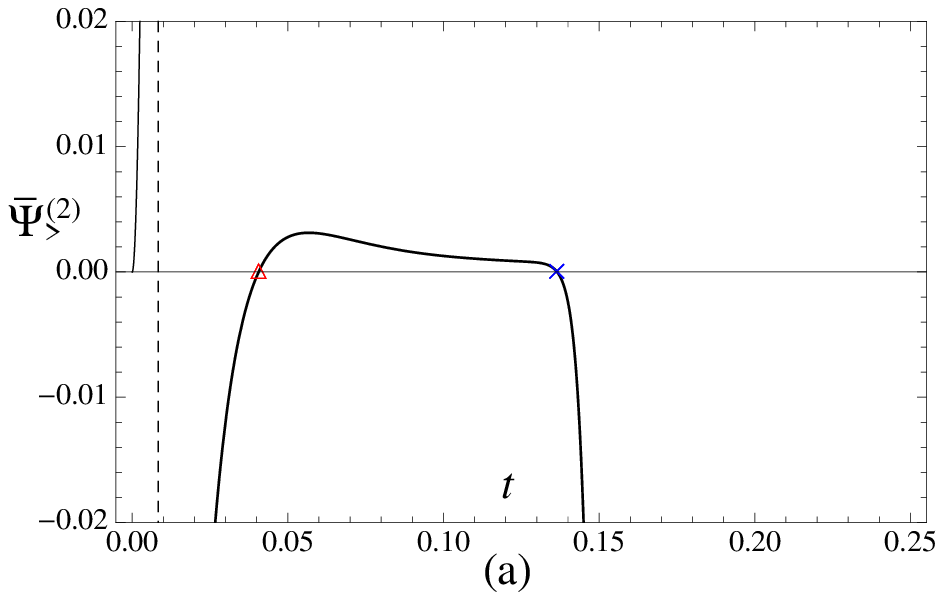}
\includegraphics[scale=0.8]{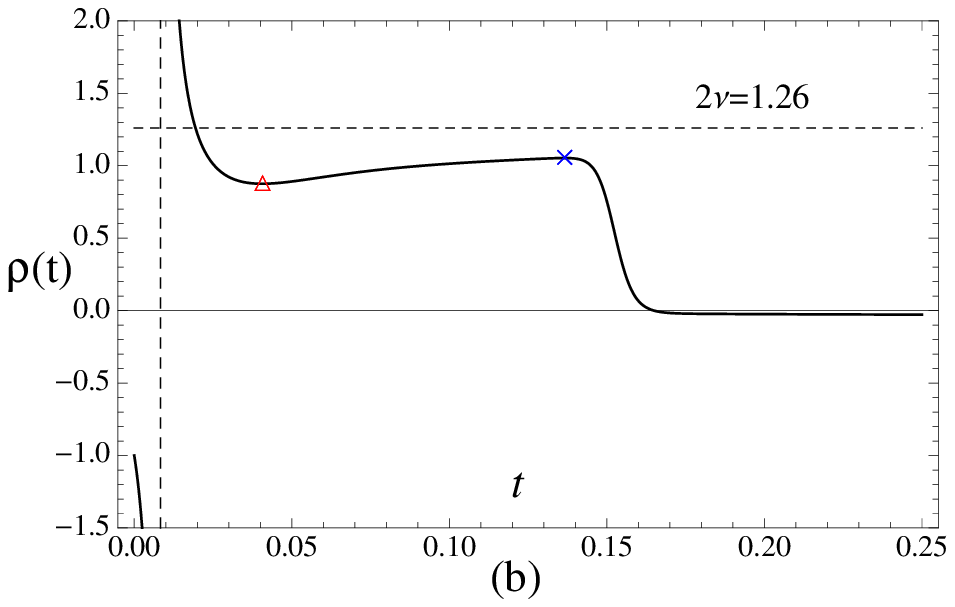}
\caption{(Color online) Plots of (a) $\bar\Psi_{>}^{(2)}|_{\rho(t)}=((\bar\beta_{>N}^{(2)})^2-\bar\beta_{>N}^{(1)}\bar\beta_{>N}^{(3)})/\bar\beta_{>N}^{(2)}$ and (b) $\rho(t)$ at $N=24$.  Scaling regions have developed broader compared to the $13$th order.  This agrees with the assumption (\ref{clue}).}
\end{figure}
Turning to the even high order case $N=24$, from the plots shown in Fig. 5, we find that the point marked by the blue cross gives the best solution where $\bar\Psi_{>}^{(2)}=0$.  

We call the solution marked by the blue cross as the proper solution and take it as giving the legitimate estimate of $\beta_{c}$.  In this manner, we can identify the proper sequence of estimated $\beta_{c}$.   In Fig. 6, the sequence of $\beta_{c}$ estimated at non-proper $t$ is shown by red triangles and at proper $t$ by blue crosses.  It is clearly shown that the proper estimates serve us better values.
\begin{table*}
\caption{Estimation of $\beta_{c}$ and $p_{1}=1/(2\nu)=0.79365$ ($1/p_{1}\sim1.26$) with one parameter.  Only the proper values are shown.}
\begin{center}
\begin{tabular}{ccccccc}
\hline\noalign{\smallskip}
$order$  & 20 & 21 & 22 &  23 & 24 & 25   \\
\noalign{\smallskip}\hline\noalign{\smallskip}
$\beta_{c}$   &  0.220933    & 0.220944    & 0.221035 &  0.221040 & 0.221114  & 0.221117\\
$1/p_{1}$   & 1.028607    &  1.029841    & 1.041745 & 1.042425 &  1.053098   & 1.053408\\
$ t^{*} $  &  0.133502    &  0.126393   &  0.135082  &  0.128043 &  0.136505  &  0.129586 \\
\noalign{\smallskip}\hline
\end{tabular}
\end{center}
\end{table*}
\begin{figure}[ht]
\centering
\includegraphics[scale=0.8]{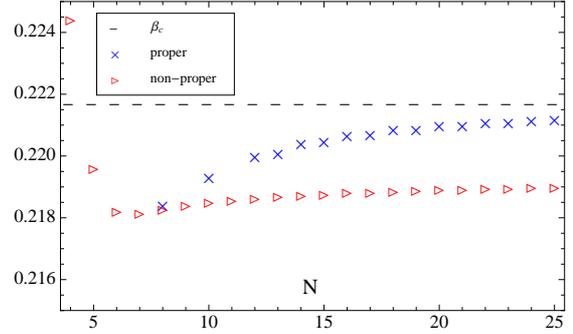}
\caption{(Color online) Estimated result of $\beta_{c}$ from $4$th to $25$th orders.  The channel of proper sequence opens at $N=8$ for even orders and $N=13$ for odd orders.}
\end{figure}

As the order increases, the optimized value of $p_{1}$ gradually increases but is not sure whether the sequence converges to the established value, $1/p_{1}=2\nu=1.2602$.  See Table 1.  On the other hand, the accuracy of estimated $\beta_{c}$ is good for $N=25$ as $\beta_{c}=0.221117\cdots$ and the relative error is about $0.24$ \%.    The non-accurate results for $p_{1}^{-1}$ may be explained as follows:  In our method of using extended PMS, the value of $p_{1}$ is fixed to achieve the total corrections being suppressed.  Here notice that the dominant part of total corrections are composed by first few or several terms in (\ref{delbetascaling}) and not only by $\bar A_{1}t^{-p_{1}}$.  Hence, even when the cancelation is successful, it does not necessarily mean that $1/p_{1}^{*}$ contains accurate information of true $1/p_{1}=2\nu$, unless the order of $1/M$ expansion is extremely large.  

The sequence of good estimations starting with $8$th order for even $N$ and $13$th for odd $N$ has enough number of terms to do extrapolation to the infinite order by a fit assuming the simplest form, $\beta_{c}(N)\sim \beta_{c}-b\times N^{-u}$. Using the last three estimations of odd orders, $21$st, $23$rd and $25$th, we obtain 
$\beta_{c}=0.221642$
with $b=0.101171$ and $u=1.634371$.  If we use the $20$th, $22$nd and $24$th orders results, we obtain $\beta_{c}=0.221624$ with $b=0.10283$ and $u=1.669628$.   Both values come to close to the established value, $\beta_{c}=0.22165$  (see (\ref{betac})).  The relative error is only $0.00035\sim 0.01 $ \%.   Also for $p_{1}^{-1}$ we have done the fitting and obtained with $21$st, $23$rd and $25$th order results, 
$p_{1}^{-1}\approx 1.2828$ and with $20$th, $22$nd and $24$th orders, $p_{1}^{-1}\approx 1.2640$.  
The estimation is much improved as having relative discrepancy to $1.2602$ about $0.3\sim 1.8$ \%.  We can say that the proper sequence tends to the present world standard.

\subsection{Estimates of $\beta_{c}$ and $\nu$ with the 2nd order LDE}
In this subsection, we apply our method to the ansatz $\beta_{<}=\beta_{c}+const\times t^{-p_{1}}+const\times t^{-p_{2}}$ which satisfies the following LDE,
 \begin{equation}
\bar\Psi_{<}=\Big[1+p_{2}^{-1}\frac{d}{d\log t}\Big]\Big[1+p_{1}^{-1}\frac{d}{d\log t}\Big]\bar \beta_{<N}=\beta_{c}.
\label{delpsilimit_multi}
\end{equation}
By using $\bar\beta_{>N}$ and its derivatives in the place of $\bar\beta_{<N}^{(k)}$, we must search three optimal values of $p_{1}$, $p_{2}$ and $t$.  For the purpose, we use extended PMS specified by the couple of equations,
\begin{eqnarray}
\bar\Psi_{>}^{(1)}=\bar\beta_{>N}^{(1)}+\rho \bar\beta_{>N}^{(2)}+\sigma \bar\beta_{>N}^{(3)}&=&0,\nonumber\\
\bar\Psi_{>}^{(2)}=\bar\beta_{>N}^{(2)}+\rho \bar\beta_{>N}^{(3)}+\sigma \bar\beta_{>N}^{(4)}&=&0,\nonumber\\
\bar\Psi_{>}^{(3)}=\bar\beta_{>N}^{(3)}+\rho \bar\beta_{>N}^{(4)}+\sigma \bar\beta_{>N}^{(5)}&\sim &0,
\label{2para}
\end{eqnarray}
where
\begin{equation}
\rho=p_{1}^{-1}+p_{2}^{-1},\quad \sigma=(p_{1}p_{2})^{-1}.
\end{equation}

It is interesting to see the results of estimation, when the $5$th order derivative, that shows the behavior insufficient for the scaling, enters into the estimation program.  As in the case of  one-parameter ansatz, 
it proves convenient to express $\rho$ and $\sigma$ in terms of $t$.  From the first two equations, we obtain
\begin{eqnarray}
\rho(t)&=&\frac{\bar\beta_{>N}^{(2)}\bar\beta_{>N}^{(3)}-\bar\beta_{>N}^{(1)}\bar\beta_{N}^{(4)}}{\Delta},\label{rho-sol}\\
\sigma(t)&=&\frac{\bar\beta_{>N}^{(1)}\bar\beta_{>N}^{(3)}-(\bar\beta_{>N}^{(2)})^2}{\Delta},
\label{sig-sol}
\end{eqnarray}
where
\begin{equation}
\Delta=\bar\beta_{>N}^{(2)}\bar\beta_{>N}^{(4)}-(\bar\beta_{>N}^{(3)})^2.
\label{Delta}
\end{equation}
The two parameters behave at large $t$ as
\begin{eqnarray}
\rho(t)&\sim& \frac{p_{1}+p_{2}}{p_{1}p_{2}}+const\times t^{-p_{3}+p_{2}}+\cdots,\nonumber\\
\sigma(t)&\sim&\frac{1}{p_{1}p_{2}}+const\times t^{-p_{3}+p_{2}}+\cdots.
\end{eqnarray}
The critical behavior of $\bar\Psi_{>}^{(3)}=\bar\beta_{>N}^{(3)}+\rho(t) \bar\beta_{>N}^{(4)}+\sigma(t) \bar\beta_{>N}^{(5)}$ is then given by $\sim const\times t^{-p_{3}}$ and decays faster than $t^{-p_{1}}$.   It should also be noted that the coefficient of $t^{-p_{3}}$ denoted by "const" depends on the order $N$ and tends to zero in the $N\to \infty$ limit.   Therefore the $3$rd order derivative in which the solutions (\ref{rho-sol}) and (\ref{sig-sol}) are substituted should exhibit rapid trend of vanishing.  At $N=25$, this is numerically confirmed as shown in Fig. 7.  The vertical dotted lines indicate the singularities of the function.  All of them come from zero points of $\Delta$, the solutions of $\Delta=0$.   The proper stationary point lies between the two singular points.  Actually, $\bar\Psi_{>}^{(3)}$ stays close to the horizontal axis between the two marking points (red triangle and blue cross)  and the expected scaling (though not so clear) is observed there.  Thus, the proper solution comes from the point represented by the blue cross.  Now the solution of $t$ is obtained by searching zero or minimum point of $\bar\Psi_{>}^{(3)}=\bar\beta_{>N}^{(3)}+\rho(t) \bar\beta_{>N}^{(4)}+\sigma(t) \bar\beta_{>N}^{(5)}$.  Then, with obtained $t^{*}$, $\rho^{*}=\rho(t^{*})$ and $\sigma^{*}=\sigma(t^{*})$, 
the estimation of $\beta_{c}$ is directly given by
\begin{equation}
\beta_{c}=\bar\Psi_{>}\Big |_{t^{*}}=\bar\beta_{>N}+\rho^{*} \bar\beta_{>N}^{(1)}+\sigma^{*} \bar\beta_{>N}^{(2)}\Big |_{t^{*}}.
\end{equation}
\begin{figure}
\centering
\includegraphics[scale=0.8]{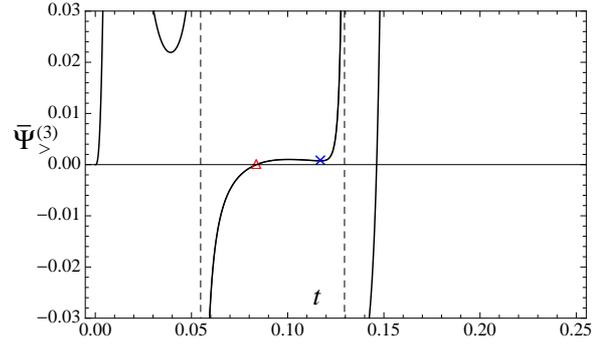}
\caption{(Color online) $\bar\Psi_{>}^{(3)}(\rho(t),\sigma(t),t)$ at $N=25$.  In the plot, we emphasized two candidates both of which provide positive two parameters $\rho$ and $\sigma$.  The blue cross at the local minimum point indicates the proper solution which is considered as lying in the scaling region.  The red triangle indicates the other, non-proper solution.}
\end{figure}
\begin{table*}
\caption{Estimation of $\beta_{c}$, $p_{1}$ and $p_{2}$ with two parameters.}
\begin{center}
\begin{tabular}{ccccccc}
\hline\noalign{\smallskip}
$order$ & 20 & 21 & 22 & 23 & 24 & 25   \\
\noalign{\smallskip}\hline\noalign{\smallskip}
$\beta_{c}$ &  0.221442  & 0.22140 & 0.221505 &   0.221513  &  0.221553  & 0.221562 \\
$1/p_{1}$ & 1.138882 & 1.127291 & 1.160289  & 1.163069  &  1.179823  & 1.183677   \\
$1/p_{2}$ & 0.433345 & 0.413636  & 0.468705  & 0.472926  & 0.500057 &  0.505965\\
$ t^{*} $  & 0.116906 &  0.096589  & 0.119071 &  0.114190  &  0.120643  &  0.116991 \\
\noalign{\smallskip}\hline
\end{tabular}
\end{center}
\end{table*}

\begin{figure}[h]
\centering
\includegraphics[scale=0.8]{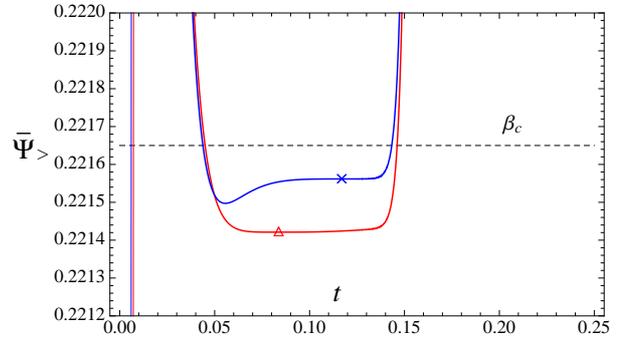}
\caption{(Color online) $\bar\Psi_{>}(\rho^{*}, \sigma^{*}, t)$ for $\rho^{*}=1.6896423$, $\sigma^{*}=0.5988992$ (blue curve) and for $\rho^{*}=1.5529395$, $\sigma^{*}=0.4763825$ (red curve).  The dashed line indicates $\beta_{c}=0.22165$.}
\end{figure}

\begin{figure}[h]
\centering
\includegraphics[scale=0.8]{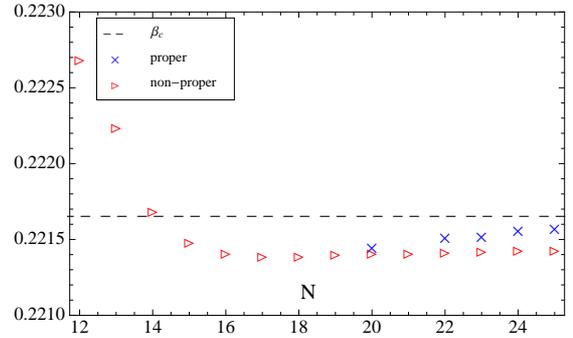}
\caption{(Color online) Plots of estimated $\beta_{c}$ with two-parameter ansatz.  The dashed line indicates $\beta_{c}=0.22165$.  The blue cross indicates proper estimate which appears from $20$th order.}
\end{figure}
The result of estimation is shown in Table 2, Figs. 8 and 9.   The behavior of $\bar\Psi_{>}^{(3)}$ becomes steady at $12$th order and single solution is obtained up to $19$th order.  The new channel to the proper sequence opens from $20$th for even orders and from $23$rd for odd orders.  The level of the scaling behavior observed in $\bar\Psi_{>}^{(3)}$ is lower compared to the one-parameter ansatz.  For example, the scaling level of $25$th order in the two-parameter case is, to the eye, the same level with the $13$th order in the one-parameter case.   However, we notice that the accuracy is improved:  The relative error of $25$th order $\beta_{c}$ is about $0.004$ \%.  The estimation of $p_{1}^{-1}=2\nu$ is not so good yet, though two-parameter estimation is improved compared to one-parameter case.  At $25$th order, the relative error is about $6$ \%.

The extrapolation of the proper sequence to the $N\to \infty$ limit is not adequate in the two-parameter case.  This is because the number of elements in the sequence is not enough.

\subsection{Estimates of $\beta_{c}$ and $\nu$ with the $3$rd and higher order LDEs}
The three parameter 
ansatz takes the form $\beta_{<}=\beta_{c}+const\times t^{-p_{1}}+const\times t^{-p_{2}}+const\times t^{-p_{3}}$.  With third order LDE, 
\begin{equation}
\bar\Psi_{<}=\prod_{i=1}^{3}\Big[1+p_{i}^{-1}\frac{d}{dt}\Big]\bar \beta_{<N}\sim \beta_{c},
\end{equation}
we continue estimation of $\beta_{c}$ and $\nu$.  
The estimation procedure follows those of 1st and 2nd order LDEs and we omit the details and present just the outline.  The extended PMS condition reads
\begin{eqnarray}
\bar\Psi_{>}^{(1)}=\bar\beta_{>N}^{(1)}+\rho \bar\beta_{>N}^{(2)}+\sigma \bar\beta_{>N}^{(3)}+\tau \bar\beta_{>N}^{(4)}&=&0,\nonumber\\
\bar\Psi_{>}^{(2)}=\bar\beta_{>N}^{(2)}+\rho \bar\beta_{>N}^{(3)}+\sigma \bar\beta_{>N}^{(4)}+\tau \bar\beta_{>N}^{(5)}&=&0,\nonumber\\
\bar\Psi_{>}^{(3)}=\bar\beta_{>N}^{(3)}+\rho \bar\beta_{>N}^{(4)}+\sigma \bar\beta_{>N}^{(5)}+\tau \bar\beta_{>N}^{(6)}&=&0,\nonumber\\
\bar\Psi_{>}^{(4)}=\bar\beta_{>N}^{(4)}+\rho \bar\beta_{>N}^{(5)}+\sigma \bar\beta_{>N}^{(6)}+\tau \bar\beta_{>N}^{(7)}&\sim &0,
\label{3para}
\end{eqnarray} 
where $\rho$, $\sigma$ and $\tau$ are parametrised as
\begin{eqnarray}
\rho&=&\frac{1}{p_{1}}+\frac{1}{p_{2}}+\frac{1}{p_{3}},\nonumber\\
\sigma&=&\frac{1}{p_{1}p_{2}}+\frac{1}{p_{2}p_{3}}+\frac{1}{p_{3}p_{1}},\nonumber\\
\tau&=&\frac{1}{p_{1}p_{2}p_{3}}.
\end{eqnarray}
By the first three equations of (\ref{3para}), the three parameters $\rho$, $\sigma$ and $\tau$ are obtained as functions of $t$ and then the substitution of them into the last equation of (\ref{3para}) gives the solution $t=t^{*}$.  Then $\rho$, $\sigma$ and $\tau$ are estimated by $\rho^{*}=\rho(t^*)$, $\sigma^{*}=\sigma(t^*)$ and $\tau^{*}=\tau(t^*)$ and $\beta_{c}$ is obtained by $\beta_{c}=\bar\beta_{>N}+\rho^{*}\bar\beta_{>N}^{(1)}+\sigma^{*}\bar\beta_{>N}^{(2)}+\tau^{*}\bar\beta_{>N}^{(3)}|_{t=t^{*}}$.

To $22$nd order, the fourth order derivative $\bar\Psi_{>}^{(4)}$ shows a complicated behavior.   Only from $23$rd order, our method begins to provide estimations characteristic to the 3-parameter ansatz.  The behavior of $\bar\Psi_{>}^{(4)}$ is not developed yet compared to the higher order results in the 1- and 2-parameter cases.  It is typically reflected to the narrowness of the region ($\rho$, $\sigma$ and $\tau$ are positive there) where the scaling is expected to emerge there at higher orders.  In addition, at the last even order $N=24$, the behavior of $\bar\Psi_{>}^{(4)}$ resembles to the $12$th or $14$th orders in the two-parameter ansatz.  See the plot (a) in Fig. 10.  These features are understandable since three-parameter ansatz involves $7$th order derivative (see (\ref{3para})) which shows the behavior far from the scaling to $25$th order.  
\begin{figure}[h]
\centering
\includegraphics[scale=0.8]{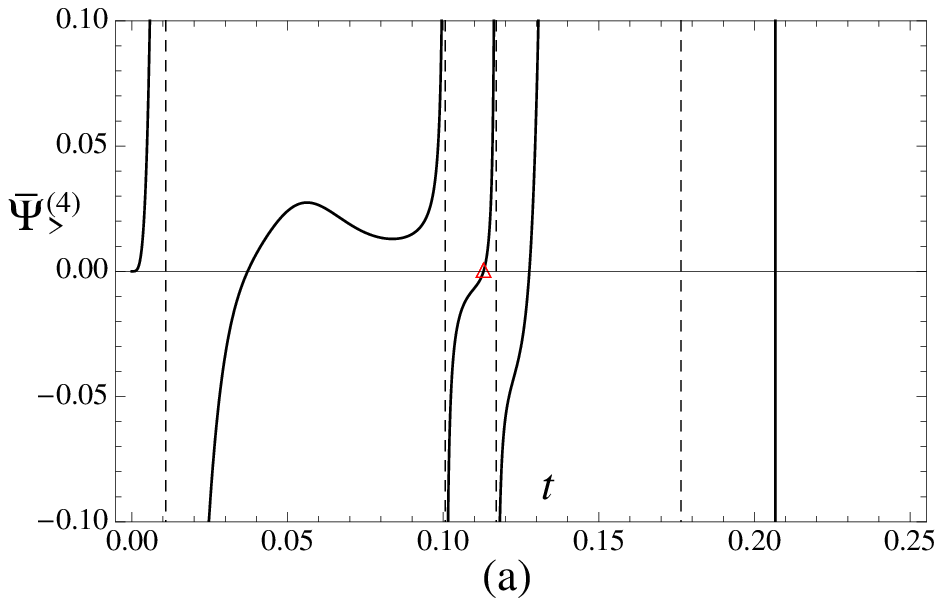}
\includegraphics[scale=0.8]{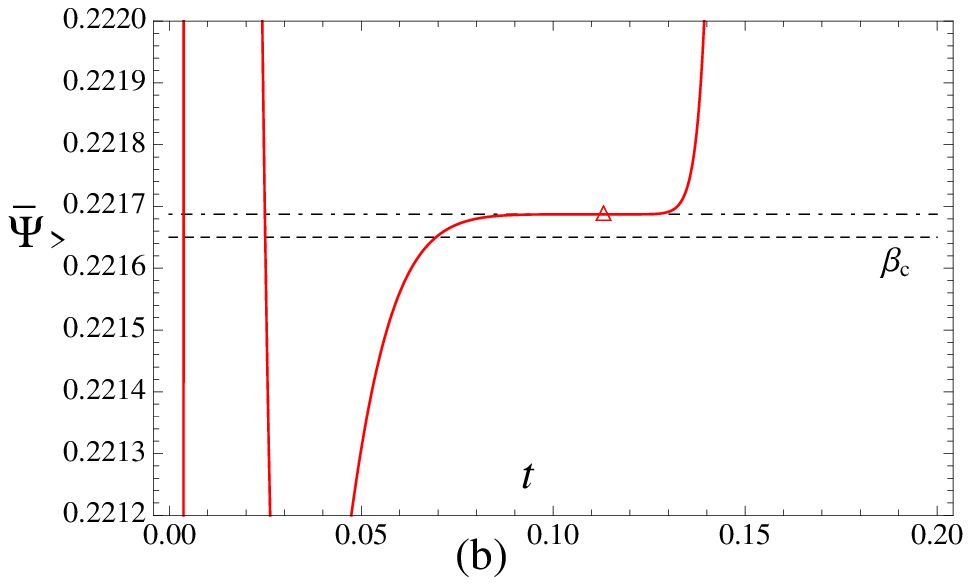}
\caption{(Color online)  (a): $\bar\Psi_{>}^{(4)}(t)$ at $N=25$.  The solution of $\bar\Psi_{>}^{(4)}=0$ indicated by the red triangle, $t^{*}=0.1129729$, is the only one (though not the proper one yet) to be considered as representing the legitimate estimation.  The solution gives $\rho^{*}=2.1339308$, $\sigma^{*}=1.2321906$ and $\tau^{*}=0.1743783$.  (b):  $\bar\Psi_{>}(\rho^{*},\sigma^{*},\tau^{*}, t)$ is plotted at $N=25$.  The red triangle gives estimate of $\beta_{c}=0.22165$, which is indicated by the dashed line.}
\end{figure}

Table 3 shows the results at $23$rd, $24$th and $25$th orders.   These three estimations exceed $\beta_{c}=0.22166$, though the last order estimation is most close to the established value.  We notice that at $25$th order, the estimation of $p_{1}^{-1}$ is also most accurate to give $\nu=0.6372$.  This implies that when $\beta_{c}$ is precisely obtained, estimated $\nu$ is also accurate.   The opening of the proper and accurate estimate for the 3rd order LDE case demands further computation of high temperature expansion, maybe up to $30$th order or more.   

\begin{table}[b]
\caption{Estimation of $\beta_{c}$ and $p_{1}$ and $p_{2}$ with three parameters.}
\begin{center}
\begin{tabular}{cccc}
\hline\noalign{\smallskip}
$order$ &  23 &  24 & 25    \\
\noalign{\smallskip}\hline\noalign{\smallskip}
$\beta_{c}$  & 0.222079  & 0.221741 & 0.221687   \\
$1/p_{1}$ & 2.010008 & 1.336976 & 1.274452   \\
$1/p_{2}$ &0.920863 & 0.715954  & 0.648486   \\
$1/p_{3}$ &0.299087 & 0.242590  & 0.210993  \\
$ t^{*} $  & 0.111805 & 0.111116  & 0.112973  \\
\noalign{\smallskip}\hline
\end{tabular}
\end{center}
\end{table}

Under the simple estimation without using extrapolation to the infinite order, increasing the number of exponents incorporated improves the estimation of critical quantities so far.  However, the reliable estimation with confidence of the scaling behavior of relevant functions sets in at larger orders when the number of exponents in the ansatz are increased.  This stems from the fact that the differentiation on $\bar\beta_{N}(t)$ creates oscillation and delays the appearance of the scaling behavior.  Let us briefly argue the points:  The small $t$ behavior of $\bar\beta_{<N}^{(k)}(t)$ is given by $\bar\beta_{<}^{(k)}(x)=(-1)^{k+1}(p_{1}^{k}\bar A_{1}t^{-p_{1}}+p_{2}^k \bar A_{2}t^{-p_{2}}+\cdots)$.  From $0<p_{1}<p_{2}<p_{3}<\cdots$ and the result in Table 3, we find $p_{2}>1$.  Then for $n\ge 2$,  $p_{n}^k$ $(n\ge 2)$ grows with $k$.   This means that the differentiation enhances higher order corrections and the critical behavior of $\bar\beta_{<}^{(k)}$ is obscured.   While at small $t$, the differentiation on $\bar\beta_{>}$ creates $n$ to the coefficient of $t^{n}$.  Hence, also in small-$t$ expansion, the upper limit of effective region of $\bar\beta_{>N}^{(k)}$ tends to shrink.  We have actually found that, up to $25$th order, $m$-parameter extension when $m\ge 4$ does not work (In the 4-exponents ansatz, $\beta^{(9)}$ enters into the estimation task and the function does not show any sign of the scaling to $25$th order).    
In our study up to $25$th order, the 3rd order LDE with three exponents is at the limit of our method.

It would be better to mention on the $p_{2}$ estimation.  As would be understood from Table 2 and Table 3, the estimation of $p_{2}$ is not successful yet.  If one uses standard values for $\nu=0.6301$ and $\theta=0.5$, one has $p_{2}^{-1}=2\nu/(1+\theta)\sim 0.84$.  Our estimated results in two- and three- parameters ansatze are still far from the value.  In the following section, we turn to the improved estimation on the critical exponents including the correction to the scaling $\omega=\theta/\nu$.

\section{Improved estimation of critical exponents}
We here attempt to improve estimations of critical exponents.  The reference values are (\ref{nu_s}) and the following summarized in ref. \cite{peli},
\begin{eqnarray}
\omega&=&0.84(4),\label{omega_s}\\
\eta&=&0.0364(5),\label{eta_s}\\
\gamma&=&1.2372(5)\label{gamma_s}.
\end{eqnarray}
The number in the parenthesis indicates the uncertainty or error in the last digit.   Recent works added further estimations; $\omega=0.782(5)$, $\eta=0.0318(3)$, $\gamma=1.2411(6)$ from \cite{pog}, $\omega=0.832(6)$, $\eta=0.03627(10)$ from \cite{has}, 
$\omega=0.82(4)$, $\eta=0.034(5)$ from \cite{lit}, $\omega=0.80(1)$, $\eta=0.0341(5)$ from \cite{gli}, 
$\omega=0.8303(18)$, $\eta=0.03631(3)$ from \cite{she} and $\omega=0.83(9)$ from \cite{harada}.  Though some results are out of the range indicated in (\ref{omega_s}) to (\ref{gamma_s}), we use those values quoted in \cite{peli} as the bench mark.

\subsection{Preliminary studies}
From the lesson of the previous section, we learned that the massless limit of a given function itself is approximated better than the exponents of the corrections included.  Thus, we use functions derived from $\beta(M)$ and $\chi(M)$ which provide associated exponents as the leading terms in the $M\to 0$ limit.  

To improve the estimation of the critical exponents, we will use the characteristic structure of 
(\ref{tauscale}), (\ref{beta_scaling}) and LDE, leading self-consistent point of view for $\nu$.  To prepare the estimation work, we study the expansion structure of relevant functions appropriate for the estimation of critical exponents.

For the estimation of $\nu$, the ratio $\beta^{(2)}/\beta^{(1)}=:f_{\beta}$ is convenient since $\nu$ itself appears as the leading term,
\begin{equation}
\frac{\beta_{<}^{(2)}}{\beta_{<}^{(1)}}=f_{\beta<}=-\frac{1}{2\nu}+B_{1} x^{-\frac{\theta}{2\nu}}+B_{2}x^{-\frac{1}{2\nu}}+\cdots.
\label{ratio3d}
\end{equation}
The amplitudes $B_{1}$, $B_{2}$, $\cdots$ are written by $a$, $\nu$, $\theta$ and so on (see (\ref{xi})) but details are not relevant for our purpose.   

The magnetic susceptibility $\chi$ is defined by
\begin{equation}
\chi=\sum_{n:sites}< s_{0}s_{n}>.
\end{equation}
To obtain $\chi$ in the critical region expressed in terms of $x$, it suffices to substitute $\tau(x)$ in (\ref{tauscale}) into the standard expression of $\chi_{<}$, $\chi_{<}(\tau)\sim C\tau^{-\gamma}[1+const(\tau^{\theta}+\cdots)+\cdots+const(\tau+\cdots)]+\chi_{R}$ ($\chi_{R}$ denotes the analytic back ground) \cite{aha}.  The result reads
\begin{equation}
\chi_{<}=Cx^{\frac{\gamma}{2\nu}}(1+const\cdot x^{-\frac{\theta}{2\nu}}+const\cdot x^{-\frac{1}{2\nu}}+\cdots).
\end{equation}
Scaling relation due to Fisher \cite{fish} reads that $\gamma/(2\nu)=1-\eta/2$ and what we can directly estimate is $\eta$ rather than $\gamma$.   In the estimation of $\eta$, it proves convenient to address $(\log\chi)^{(1)}=(d/d\log x)\log \chi:=f_{\chi}$.   It behaves near the critical point,
\begin{equation}
f_{\chi<}=\frac{\gamma}{2\nu}+C_{1} x^{-\theta/2\nu}+C_{2} x^{-1/2\nu}+\cdots.
\label{chiscaling}
\end{equation}

All behaviors of the series (\ref{ratio3d}) and (\ref{chiscaling}) can be written as
\begin{equation}
f_{<}=f_{c}+f_{1}x^{-q_{1}}+f_{2}x^{-q_{2}}+\cdots,
\label{general_scaling}
\end{equation}
where
\begin{equation}
q_{1}=\frac{\omega}{2},\quad q_{2}=\frac{1}{2\nu}.
\end{equation}
The function $f_{<}$ satisfies $K$th order LDE,
\begin{equation}
\Phi_{<}:=\prod_{n=1}^{K}\Big[1+q_{n}^{-1}\frac{d}{d\log x}\Big]f_{<}=f_{c}+O(x^{-q_{K+1}}).
\label{basiclde}
\end{equation}
Hence, after the transformation by $\delta$-expansion and truncating the higher order corrections of order $O(t^{-q_{K+1}})$, we find
\begin{equation}
\bar\Phi_{<}:=\prod_{n=1}^{K}\Big[1+q_{n}^{-1}\frac{d}{d\log t}\Big]\bar f_{<N}=f_{c}.
\label{basiclde2}
\end{equation}
Our estimation protocol is almost the same as that for $\beta_{c}$.  In the place of $\bar f_{<N}$, $\bar f_{>N}$ is substituted because there is the matching region where the scaling behavior may be observed in $\bar f_{>N}$, and then $f_{c}$ and unknown critical exponent $q_{i}$ will be estimated by utilizing extended PMS and self-consistent conditions (to be clarified later).   

From $f_{\beta>}=\beta^{(2)}_{>}/\beta^{(1)}_{>}$, 
\begin{equation}
f_{\beta>}=1-12 x+104 x^2-1008 x^3+10416 x^4-\cdots,
\label{beta_high}
\end{equation}
and the $\delta$-expansion at the expansion order $N$ transforms the above result to $\bar f_{\beta >N}:=D_{N}[f_{\beta >N}]$,
\begin{equation}
\bar f_{\beta >N}=1-12 {N \choose 1}t+104{N \choose 2} t^2-1008 {N \choose 3} t^3+\cdots,
\label{beta_high}
\end{equation}
where the last term should be of the order $t^N$.  We note that, while the highest order of $\beta_{>N}$ is $25$th, the highest order of $f_{\beta>N}$ is $24$th, because of the cancellation of $t$ in the numerator and the denominator in $\beta^{(2)}_{>N}/\beta^{(1)}_{>N}$.

The susceptibility at high temperature is also written with $1/M=x\, (\ll 1)$.  First remind that $\chi_{>}=1+6\beta+30\beta^2+O(\beta^3)$ \cite{butera}.   Then, by the substitution of $\beta_{>}(x)$ given in (\ref{beta_M}) into $\beta$, we obtain
\begin{eqnarray}
\chi_{>}&=&1+6x-6x^2+36x^3-270x^4+2268x^5\nonumber\\
& &-20436x^6+193176x^7-1890462x^8+18990892x^9\nonumber\\
& &-194709708x^{10}+2029271688x^{11}\nonumber\\
& &-21435300372x^{12}
+228983179752x^{13}\nonumber\\
& &-2469626018184x^{14}
+26855777435248x^{15}\nonumber\\
& &-294145354348974x^{16}+3242105906258220x^{17}\nonumber\\
& &-35935261094616124x^{18}\nonumber\\
& &
+400295059578038760x^{19}\nonumber\\
& &-4479014443566807276x^{20}\nonumber\\
& &+50319506857313420376x^{21}\nonumber\\
& &-567383767790459777016x^{22}\nonumber\\
& &+6418899321986117552400x^{23}\nonumber\\
& &-72838651914163555355012x^{24}\nonumber\\
& &+828839976149614386374184x^{25}-\cdots,
\end{eqnarray}
and from $f_{\chi>}=\chi^{(1)}_{>}/\chi_{>}$,
\begin{equation}
f_{\chi>}=6x-48x^2+432x^3-4167 x^4+42336 x^5-\cdots.
\label{chi_high}
\end{equation}
We thus arrive at
\begin{equation}
\bar f_{\chi >N}=6{N \choose 1}x-48 {N \choose 2}x^2+432 {N \choose 3}x^3-\cdots.
\label{chi_high}
\end{equation}
where $D_{N}[f_{\chi >N}]:=\bar f_{\chi >N}$.  
For the sake of notational simplicity, let us omit the subscript $N$ in $\bar f_{\beta(\chi) >(<)N}$ henceforth.

\subsection{Biased estimation of $\nu$}
First of all, we like to confirm that our approach provides results consistent with (\ref{nu_s}), (\ref{eta_s}) and (\ref{gamma_s}) by the use of the value $\omega=0.84(4)$ in (\ref{omega_s}), giving
\begin{equation}
q_{1}=\frac{\omega}{2}=0.42(2).
\end{equation}
Such a biased estimation strengthen the validity of the approach when successful and suggests possible improvement of the protocol in an unbiased estimation which will be explored later.

\begin{figure}
\centering
\includegraphics[scale=0.8]{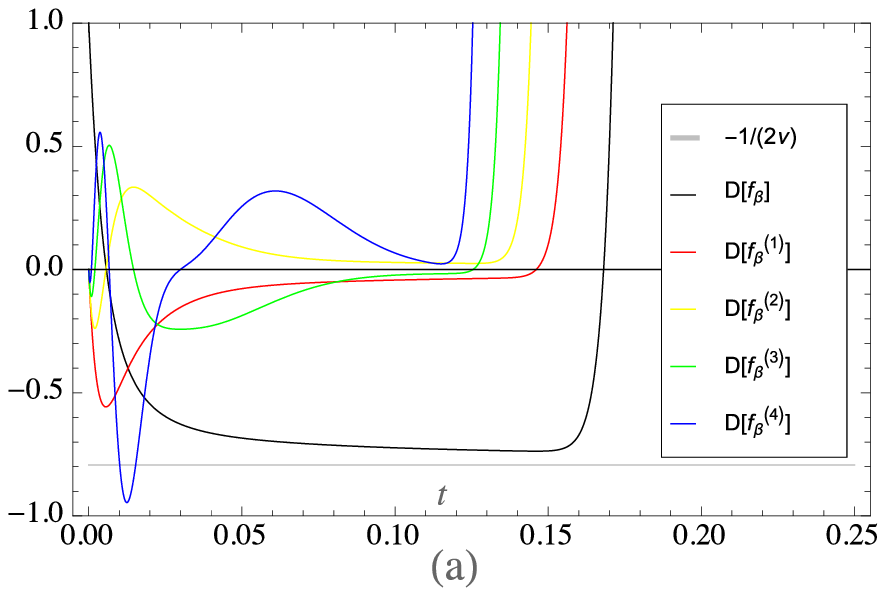}
\includegraphics[scale=0.8]{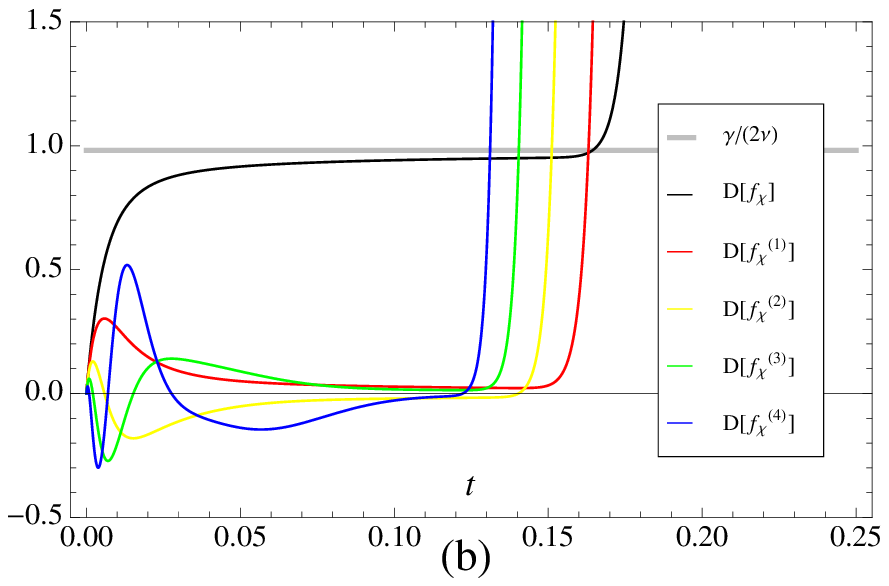}
\caption{(Color online) The graphs (a) and (b) show plots of $\bar f_{\beta>}$ and $\bar f_{\chi>}$ at respective highest orders ($N=24$ for $\bar f_{\beta>}$ and $25$ for $\bar f_{\chi>}$) and their derivatives with respect to $d/d\log t$ to the 4th order.  The gray lines indicate $-1/(2\nu)$ for $\nu=0.6301$ in (a) and $\gamma/(2\nu)$ for $\gamma=1.2373$ and $\nu=0.6301$ in (b).}
\end{figure}
\begin{figure}
\centering
\includegraphics[scale=0.8]{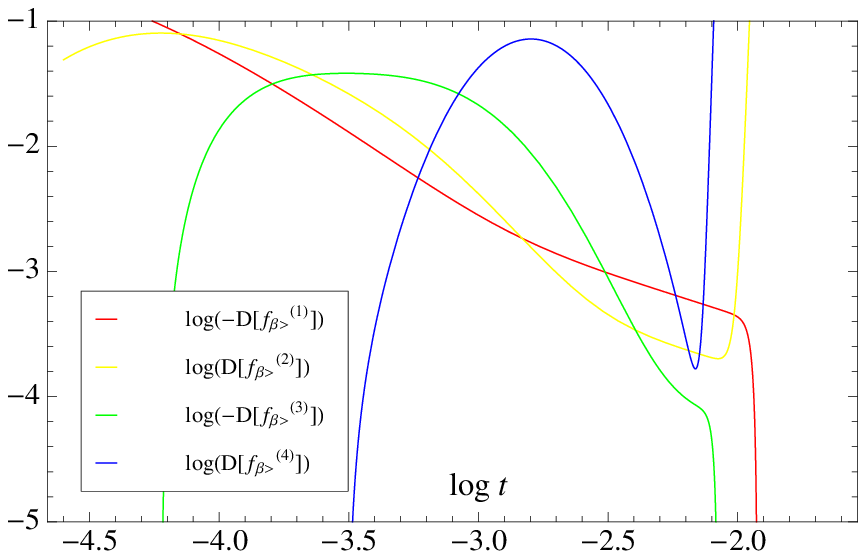}
\includegraphics[scale=0.8]{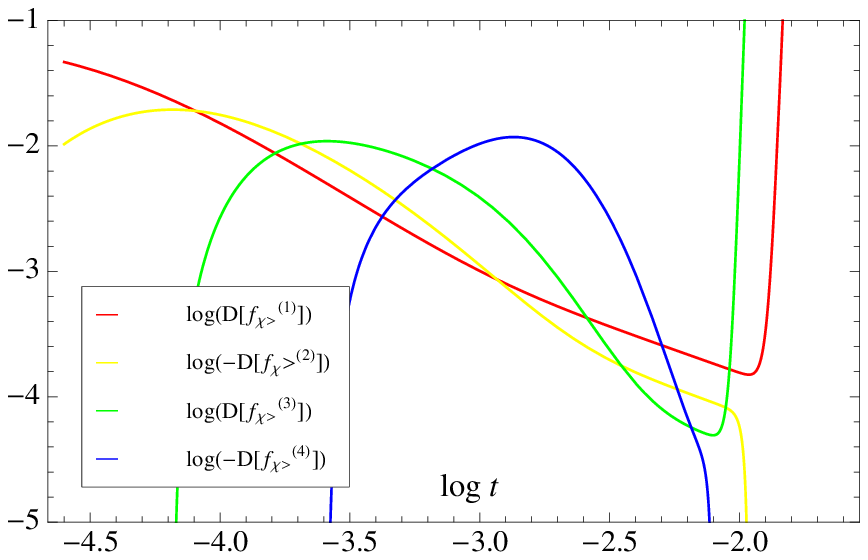}
\caption{(Color online) The graphs (a) and (b) show plots of $\log((-1)^{k}\bar f_{\beta>}^{(k)})$ and $\log((-1)^{k+1}\bar f_{\chi>}^{(k)})$ at respective highest orders ($N=24$ for $\bar f_{\beta>}$ and $25$ for $\bar f_{\chi>}$) for $k=1,2,3,4$.}
\end{figure}

Before the estimation, we survey the behaviors of $\bar f_{\beta>}$, $\bar f_{\chi>}$ and their derivatives.  We find from Fig. 11 (a), (b) that the functions to the third order derivatives  may exhibit scalings.  The plots of $\log((-1)^{k}\bar f_{\beta>}^{(k)})$ and $\log((-1)^{k+1}\bar f_{\chi>}^{(k)})$ shown in Fig. 12(a), (b) provide us more detailed information on the scaling levels. Both for $\bar f_{\beta>}$ and $\bar f_{\chi>}$, the derivatives to the second order show scalings while the third order derivative is not sufficient.  As for the fourth order one, we observe no linear-like behavior.  This implies that the derivatives to the second or at most  third order are appropriate for the estimation.  The level of the scaling shown in the derivatives is less-qualified to that of the square model (see Fig. 3(a)).

As the first protocol (i), we consider the estimation of $\nu$ via the function $f_{\beta >}$ only.  
When the first order ansatz $f_{\beta<}=-1/(2\nu)+const\times x^{-q_{1}}$ is used, the LDE to be considered is $\bar\Phi_{\beta<}=[1+q_{1}^{-1}(d/d\log t)]\bar f_{\beta<}=-1/(2\nu)$.  Then, substituting $q_{1}=\omega/2=0.84(4)/2$, we search the stationary point around $t=0.11\sim 0.13$ where the scalings of $\bar f_{\beta >}$ and $\bar f_{\beta >}^{(1)}$ are observed (see the plot of $\bar f_{\beta>}^{(k)}$ $(k=0,1)$ in Fig. 11(a).)  Then the stationary point at each order of $N=21$ to $24$ gives $\nu=0.6078(34), 0.6077(35), 0.6095(33),0.6094(33)$, respectively but these are not good approximation.  Satisfactory estimation comes from the second order LDE, where the second order ansatz reads $f_{\beta<}=-1/(2\nu)+const\times x^{-q_{1}}+const\times x^{-q_{2}}$.  The second order LDE is given by
\begin{equation}
\bar\Phi_{\beta<}=\Big[1+q_{1}^{-1}\frac{d}{d\log t}\Big]\Big[1+q_{2}^{-1}\frac{d}{d\log t}\Big]\bar f_{\beta<}=-\frac{1}{2\nu}.
\label{beta_lde1}
\end{equation}
Since $1/(2\nu)$ in the right-hand-side is nothing but $q_{2}$, we can write the above LDE as
\begin{equation}
\bar\Phi_{\beta<}=\bar f_{\beta<}+(q_{1}^{-1}+q_{2}^{-1})\bar f_{\beta<}^{(1)}+(q_{1}q_{2})^{-1}\bar f_{\beta<}^{(2)}=-q_{2}.
\label{beta_lde2}
\end{equation}
By this installation of the self-consistent point of view, we can reduce the number of equations and the order of derivatives included in the estimation.  
We thus estimate $\nu$ by the both use of (\ref{beta_lde2}) and the PMS condition, 
\begin{equation}
\bar\Phi_{\beta<}^{(1)}=\bar f_{\beta<}^{(1)}+(q_{1}^{-1}+q_{2}^{-1})\bar f_{\beta<}^{(2)}+(q_{1}q_{2})^{-1}\bar f_{\beta<}^{(3)}=0.
\label{pms_fbeta}
\end{equation}
In the successive steps, we replace $\bar f_{\beta <}^{(k)}$ by $\bar f_{\beta >}^{(k)}$.  It is cumbersome to remark this in each and every cases, we like to write LDEs only with $\bar f_{\beta >}^{(k)}$ and $\bar f_{\chi >}^{(k)}$.  
We first solve the LDE $\bar\Phi_{\beta>}^{(1)}=0$, and obtain $q_{2}^{-1}$ involved in the correction as the function of $t$, the estimation pout.  The solution denoted by $Q_{2\beta}(t)^{-1}$ is given by
\begin{equation}
Q_{2\beta}(t)^{-1}=-\frac{\bar f_{\beta >}^{(1)}+(2/\omega)\bar f_{\beta >}^{(2)}}{\bar f_{\beta >}^{(2)}+(2/\omega)\bar f_{\beta >}^{(3)}}.
\end{equation}
Then, substituting the solution $Q_{2\beta}(t)$ into $q_{2}$ in both sides of (\ref{beta_lde2}), we look for the self-consistent point around, say, $t=0.11\sim 0.12$.  The result is summarized in Table 4 in the $\nu_{1}$ row.  Actually, at odd orders, the solution $t$ of (\ref{beta_lde2}) just appeared from $23$rd order.  For lower odd orders, two curves of $\bar\Phi_{\beta>}$ and $-Q_{2\beta}(t)$ have no common point and we have picked up approximate solution of $t$ where $\bar\Phi_{\beta>}-(-Q_{2\beta}(t))$ becomes locally minimum in the absolute value.    Also at even orders, we consider that reliable solution just  emerge from $22$nd order, since from that order the second order derivative $\bar\Phi_{\beta>}^{(2)}|_{Q_{2\beta}(t)}$ begins to show the tendency toward zero, after the peak at lower $t$, around the self-consistent point.   We see from Table 4 that the result manifests good accuracy at high orders.  We however point out one drawback contained in the approach.  That is, the scaling of $Q_{2\beta}$ which is deduced from the differentiation of (\ref{ratio3d}) that
\begin{equation}
Q_{2\beta}(t)^{-1}=\frac{1}{q_{2}}+O(t^{-q_{3}+q_{2}}),
\label{Q2-scaling}
\end{equation}
is not seen even at higher orders such as $N=23,24$.   As seen in Fig. 13 (a), (b), the curves of $-Q_{2\beta}$ show no sign of approaching to $-1/(2\nu)$ represented by the dashed line.
\begin{figure}
\centering
\includegraphics[scale=0.8]{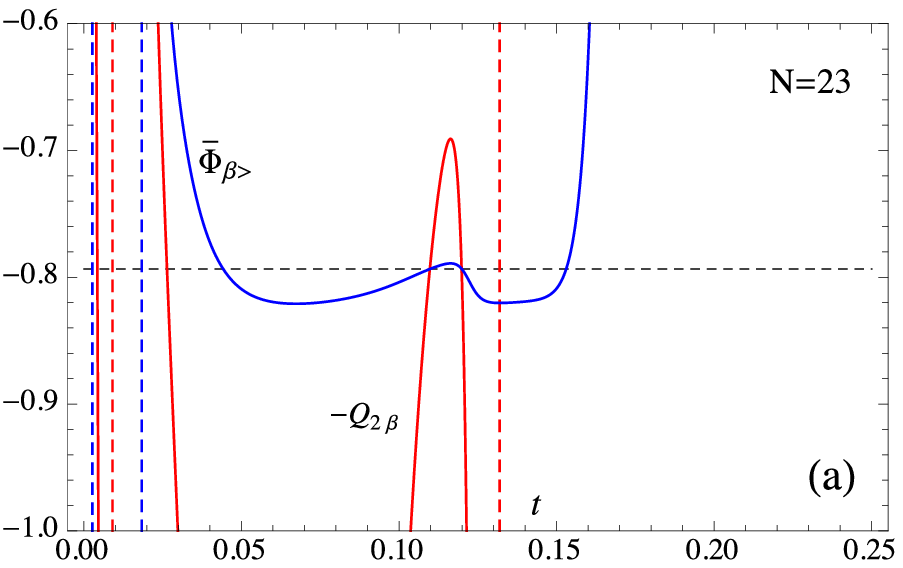}
\includegraphics[scale=0.8]{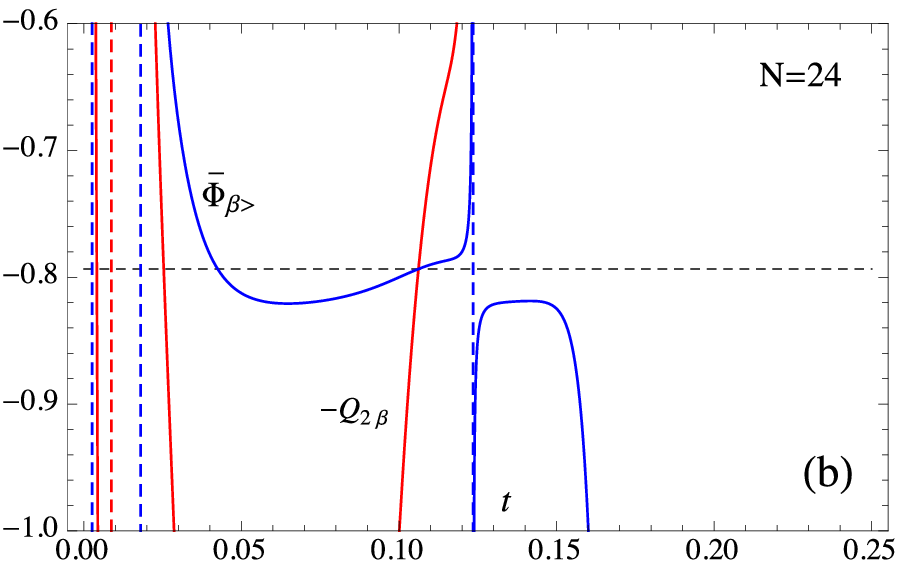}
\caption{(Color online) Plots of $\bar\Phi_{\beta>}|_{Q_{2\beta}(t)}$ (blue) at $N=23$ and $24$ and $Q_{2\beta}(t)$ (red) for $\omega=0.84$.  Horizontal dashed lines indicate the standard value of $-1/(2\nu)=-1/(2\times 0.6301)$.} 
\end{figure} 

\begin{figure}
\centering
\includegraphics[scale=0.8]{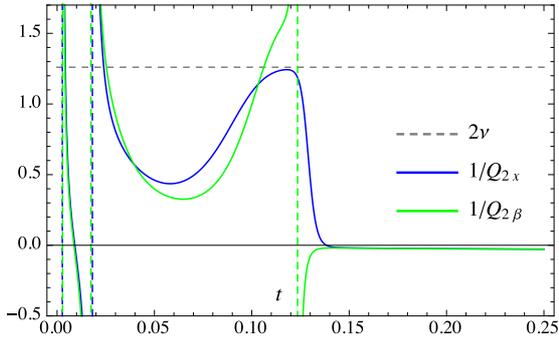}
\caption{(Color online) The graphs show plots of $Q_{2\chi}(t)$ (blue) and $Q_{2\beta}(t)$ (Green) for $\omega=0.84$ at respective highest orders ($N=25$ for the first and $24$ for the second functions).}
\end{figure}

Now, to explore the possibility of handling series for the estimation, let us consider the use of $f_{\chi}$ to replace $Q_{2\beta}$ by a better function.  The series $f_{\chi <}$ is expected to have almost the same structure with $f_{\beta<}$;  Though differences remain in the limit $x\to \infty$ and the amplitudes of each corrections, the both functions have corrections $x^{-\omega/2}$, $x^{-1/(2\nu)}$ etc.  The function $\bar f_{\chi>}$ has an advantage, compared to $\bar f_{\beta>}$, that the derivatives show better behaviors than those of $\bar f_{\beta>}$ derivatives (see Fig. 11 and Fig. 12).   For instance, from Fig. 12, we find that $\bar f_{\chi>}^{(k)}$ $(k=1,2)$ exhibits a bit longer linear-like behavior than that of $\bar f_{\beta>}^{(k)}$.   Thus, from the stationarity condition for $\bar \Phi_{\chi >}$,
\begin{equation}
\bar \Phi_{\chi >}^{(1)}=\bar f_{\chi>}^{(1)}+(q_{1}^{-1}+q_{2}^{-1})\bar f_{\chi >}^{(2)}+(q_{1}q_{2})^{-1}\bar f_{\chi >}^{(3)}=0,
\end{equation}
we obtain
\begin{equation}
Q_{2\chi}(t)^{-1}=-\frac{\bar f_{\chi >}^{(1)}+(2/\omega)\bar f_{\chi >}^{(2)}}{\bar f_{\chi >}^{(2)}+(2/\omega)\bar f_{\chi >}^{(3)}},
\label{q2chi}
\end{equation} 
which represents $q_{2}^{-1}$ at the estimation point $t$.   To demonstrate the differences between $Q_{2\beta}(t)$ and $Q_{2\chi}(t)$ clearer, we plotted them at the respective highest orders in Fig. 14.  The function $Q_{2\chi}(t)$ exhibits the rough scaling and the stationary value at the top of the hill signals $1/(2\nu)$-value (As $Q_{2\beta}$ obeys the scaling (\ref{Q2-scaling}), $Q_{2\chi}$ also obeys the same scaling.  If the scaling could be of high grade, it should develop a plateau and the stationary point indicates $q_{2}^{-1}$).   Our recipe, which is the point in this protocol (ii), is to replace $Q_{2\beta}^{-1}(t)$ by $Q_{2\chi}^{-1}(t)$ in $\bar\Phi_{\beta>}$.   Here we note that for the combined use of $\bar\Phi_{\beta >}$ and  $\bar\Phi_{\chi >}$, care must be paid that the solution of $\bar\Phi_{\chi }^{(1)}=0$ at $(N+1)$th order should be substituted into $N$th order $\bar\Phi_{\beta >}$, since both functions are derived from $(N+1)$th order expansion of $\beta(M)$.  By the substitution of $Q_{2\chi}(t)$ into $q_{2}$ in (\ref{beta_lde2}), we search the solution obeying self-consistency in exact or approximate manner.  For instance, as shown in Fig. 15 (a), the exact solution exists at $N=23$ as the intersection of two curves.  On the other hand, $N=24$ case satisfies an approximate one, $\bar\Phi_{\beta>}-(-Q_{2\chi}(t))\sim 0 (\neq 0)$ (see the behaviors around $t\sim 0.12$ plotted in Fig. 15(b)).  The result is recorded in Table 4 as a $\nu_{2}$-sequence.  As in the first protocol, the result is good at higher orders.  We add one point to be noted that in this protocol the stationarity of $\bar\Phi_{\beta>}$ at the estimation point is not exactly realized due to the replacement of $Q_{2\beta}^{-1}$ by $Q_{2\chi}^{-1}$.  However, the local gradient at the estimation point is very small and the stationarity is still respected precisely.
\begin{figure}
\centering
\includegraphics[scale=0.8]{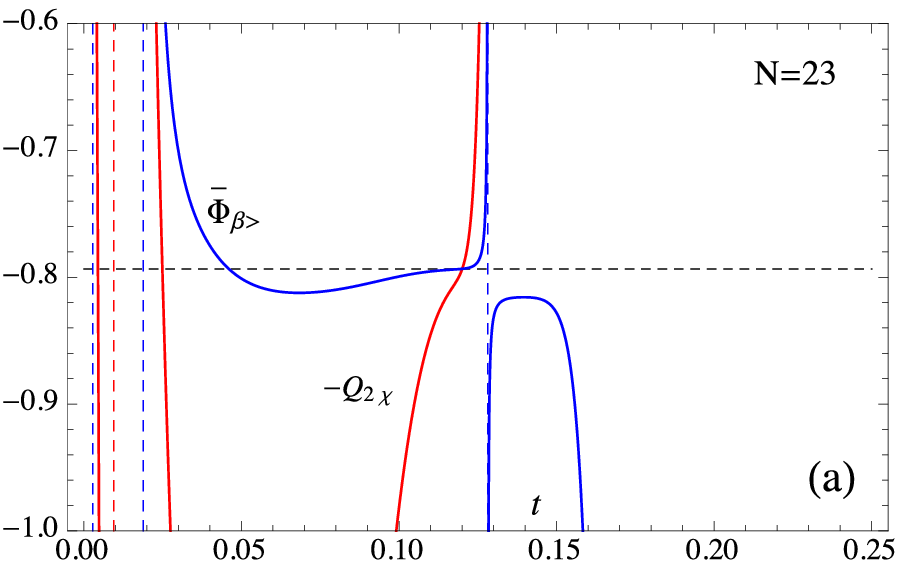}
\includegraphics[scale=0.8]{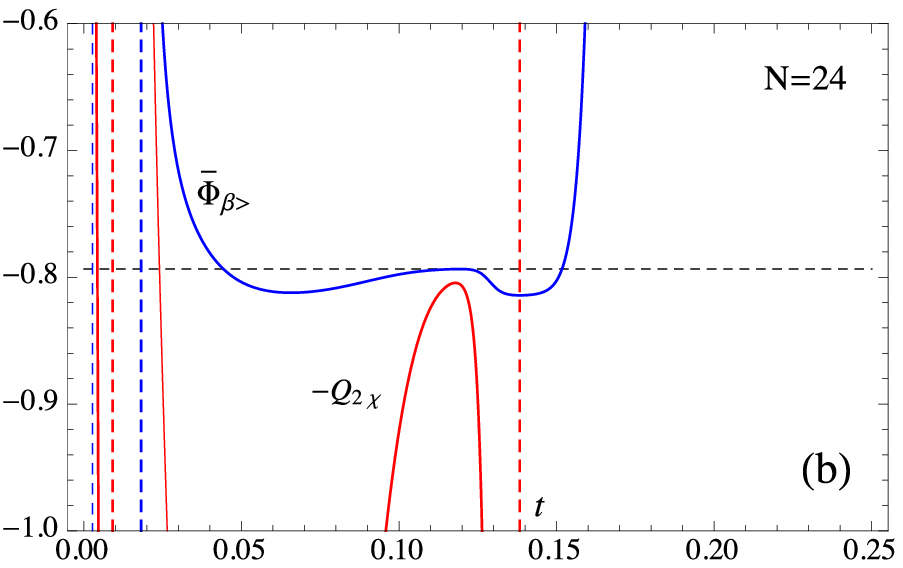}
\caption{(Color online) Plots of $\bar\Phi_{\beta>}|_{Q_{2\chi}(t)}$ (blue) at $N=23$ and $24$ and $Q_{2\chi}(t)$ (red) for $\omega=0.84$.  Horizontal dashed lines indicate the standard value of $-1/(2\nu)=-1/(2\times 0.6301)$.  In the last plot, we see that $\bar\Phi_{\beta>}|_{Q_{\chi 2}(t)}$ and $-Q_{\chi2}(t)$ come close with each others around $t\sim 0.118$ and approximately satisfies self-consistency represented in (\ref{beta_lde2}).} 
\end{figure}

As the last protocol (iii), we consider the sole use of the PMS condition in the combination of $f_{\chi}$ and $f_{\beta}$.  The recipe uses the two conditions, $\bar \Phi_{\chi >}^{(1)}=0$ and $\bar \Phi_{\beta>}^{(1)}\sim 0$.  The first condition gives $Q_{2\chi}$ (see (\ref{q2chi})) and then its substitution into $\bar \Phi_{\beta >}^{(1)}$ yields the solution   $t=t_{PMS}$ where the first derivative becomes zero or nearly zero.  Then we obtain the estimate of $\nu$ from (\ref{beta_lde1}),
\begin{equation}
\bar\Phi_{\beta>}|_{t_{PMS}}=-\frac{1}{2\nu}.
\label{nu_pms}
\end{equation}
The proper solution appeared just at $23$rd order (non-proper solution at the order, giving $\nu_{3}=0.62787$, has omitted in Table 4).  At this order, we emphasize that the self-consistency is also satisfied approximately, since $1/(2Q_{2\chi}(t_{PMS}))=0.62924$ which is close to $\nu=0.63020$ estimated by (\ref{nu_pms}).   This approximate realization of the self-consistent feature is observed for the first time at $23$rd order.   Though proper solution is not obtained at $24$th order, it would emerge at least a few larger even order.  

\begin{table*}
\caption{$\omega$-biased estimates of $\nu$ via three protocols (i), (ii) and (iii):   (i) estimates $\nu_{1}$ via $\bar \Phi_{\beta>}^{(1)}=0$ and $\Phi_{\beta>}^{(1)}=-q_{2}$.   (ii) estimates $\nu_{2}$ via $\bar \Phi_{\chi>}^{(1)}=0$ and $\Phi_{\beta>}^{(1)}=-q_{2}$.  (iii) estimates $\nu_{3}$ via the PMS conditions $\bar\Phi_{\chi }^{(1)}=0$ and $\bar\Phi_{\chi }^{(2)}=0$.  Parenthesis denotes the variation range of estimated $\nu$ under the change of $\omega$ from $0.80$ to $0.88$.}
\begin{center}
\begin{tabular}{ccccc}
\hline\noalign{\smallskip}
$order$ & 21 &  22 &  23 & 24    \\
\noalign{\smallskip}\hline\noalign{\smallskip}
$\nu_{1}$ & 0.62685(54) & 0.63004(63) & 0.63009(62)  & 0.63013(62)  \\
$\nu_{2}$  & 0.62977(62) & 0.62929(54) & 0.63023(59)  & 0.63007(47)  \\
$\nu_{3}$  & 0.62791(67) & 0.62787(76) & 0.63020(45)  & 0.62782(78)  \\
\noalign{\smallskip}\hline
\end{tabular}
\end{center}
\end{table*}
\begin{figure}
\centering
\includegraphics[scale=0.8]{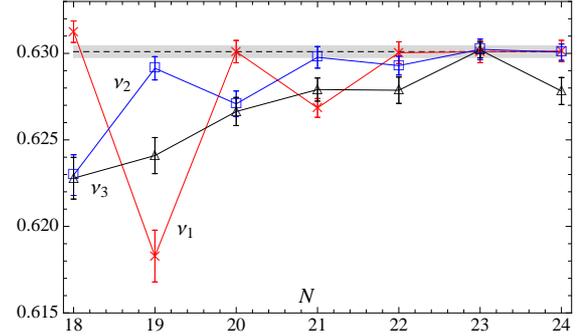}
\caption{(Color online) Plot of $\omega$-biased estimates of $\nu$ via three protocols (i), (ii) and (iii):   (i) Red plot with crosses indicates the result $\nu_{1}$ via $\bar \Phi_{\beta>}^{(1)}=0$ and $\Phi_{\beta>}^{(1)}=-q_{2}$.   (ii) Blue plot with boxes indicates the result $\nu_{2}$ via $\bar \Phi_{\chi>}^{(1)}=0$ and $\Phi_{\beta>}^{(1)}=-q_{2}$.  (iii) Black plot with triangles indicates the result $\nu_{3}$ via the PMS conditions $\bar\Phi_{\chi }^{(1)}=0$ and $\bar\Phi_{\chi }^{(2)}=0$.  Error bar denotes the variation range of estimated $\nu$ under the change of $\omega$ from $0.80$ to $0.88$ (Lower and upper ends represent the limit for $\omega=0.80$ and $0.88$, respectively).  Dashed line indicates $\nu=0.6301$ and the covered light gray zone indicates the range suggested in (\ref{nu_s}).}
\end{figure}

We have plotted $\nu_{i}\,\,(i=1,2,3)$ against the order $N$ in Fig. 16.  Though the extended PMS protocol (iii) needs a few higher orders for the assurance of convergence issue, we can say that all three protocols suggest the trend of the convergence to the correct limit.  For the protocol (iii), however, the $25$th order $1/M$ expansion ($24$th order for $f_{\beta}$) is a bit too short.  To summarize our estimation, we therefore focus on the self-consistent two protocols.  Rounding the two estimates of $\nu$ at the respective highest orders in protocols (i) and (ii) off to four decimal places, we find the both results agree with each other and yield single estimate,
\begin{equation}
\nu=0.6301(6).
\label{nu_biased}
\end{equation}
Here the number in the parenthesis implies approximate deviation when $\omega=0.80$ and $0.88$ are used.  Note that the indicated range comes from the uncertainty of $\omega$ and not from some statistical origin.  To compare our result (\ref{nu_biased}) with the reference value $\nu=0.6301(4)$ quoted in Ref. \cite{peli}, our estimate is in excellent agreement.  

\subsection{Biased estimation of $\eta$ and $\gamma$}
The exponent $\eta$ is estimated by using the first order ansatz $f_{\chi <}=\gamma/(2\nu)+const\times x^{-q_{1}}$ and the second order one, $f_{\chi <}=\gamma/(2\nu)+const\times x^{-q_{1}}+const\times x^{-q_{2}}$.  The result from the first order LDE manipulated in the same way as $f_{\beta >}$ is not acceptable, however, since the estimate gives negative value of $\eta$ in some cases.  Therefore, we need second order ansatz satisfying
\begin{equation}
\bar\Phi_{\chi >}=\bar f_{\chi >}+(q_{1}^{-1}+q_{2}^{-1})\bar f_{\chi >}^{(1)}+(q_{1}q_{2})^{-1}\bar f_{\chi >}^{(2)}=\frac{\gamma}{2\nu}.
\label{f_chi_2}
\end{equation}

There are essentially two prescriptions to estimate $\eta$.   One is, to avoid the bias as long as possible, to treat $q_{2}$ as an adjustable parameter for the PMS.  The PMS conditions to be used reads $\bar\Phi_{\chi >}^{(1)}=0$ and $\bar\Phi_{\chi >}^{(2)}\sim 0$.  From the first condition, we have $Q_{2\chi}(t)^{-1}$ in exact agreement with (\ref{q2chi}) and replace $q_{2}$ by $Q_{2\chi}(t)$ in the second derivative $\bar\Phi_{\chi >}^{(2)}$.  We then search its zero or closest point to zero in the region $t=0.11\sim 0.12$ where the emergence of the scaling is expected (See Fig. 11(b)).  By substituting the solution $Q_{2\chi}(t)$ into $\bar\Phi_{\chi>}$ we obtain the estimate of $\gamma/(2\nu)$ and then $\eta_{1}$, the estimate of $\eta$.  The other recipe is just to use estimated $q_{2}(=1/2\nu)$ obtained in the previous subsection into $\bar\Phi_{\chi >}$ and search the least variation point of $\bar\Phi_{\chi>}$ at the appropriate region.  The point provides the estimate of $\gamma/(2\nu)$ and $\eta_{2}$ of $\eta$-estimate.  As an representative value of $q_{2}$ in this prescription, we used $\nu_{2}$ in Table 4.   The result of the two estimations is summarized in Table 5 and Fig. 17.
\begin{table*}
\caption{$\omega$-biased estimation of $\eta$ with the second order LDE.  $\eta_{1}$ means the result from extended PMS and $\eta_{2}$ from the simple substitution of $\nu_{2}$ and PMS.  See the main text in detail.}
\begin{center}
\begin{tabular}{ccccc}
\hline\noalign{\smallskip}
$order$ & 22 &  23 &  24 & 25    \\
\noalign{\smallskip}\hline\noalign{\smallskip}
$\eta_{1}$ & 0.03367(61) & 0.03586(44) & 0.03672(31)  & 0.03695(25)  \\
$\eta_{2}$ & 0.03770(50) & 0.03755(48) & 0.03761(47)  & 0.03758(46)  \\
\noalign{\smallskip}\hline
\end{tabular}
\end{center}
\end{table*}
At the highest two orders, all recipes give a bit larger values compared with the standard one, (\ref{eta_s}). 
\begin{figure}
\centering
\includegraphics[scale=0.8]{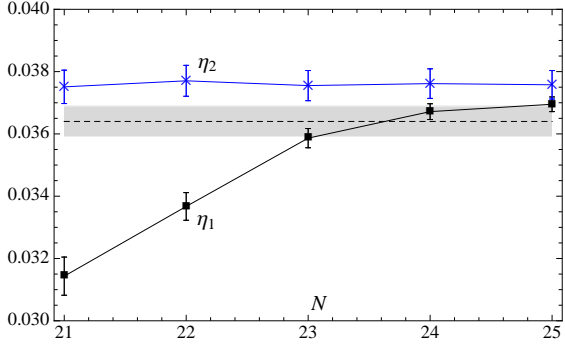}
\caption{(Color online) Plots of $\omega$-biased estimates of $\eta$ via two protocols:  Black plot indicates the results via PMS conditions $\bar\Phi_{\chi }^{(1)}=0$ and $\bar\Phi_{\chi }^{(2)}\sim 0$ without using estimated $\nu$.  Blue plot indicates the results via $\bar\Phi_{\chi }^{(1)}=0$ and with the substitution of $\nu_{2}$.  Error bar denotes the variation range of estimated $\eta$ under the change of $\omega$ from $0.80$ to $0.88$ (Lower and upper ends represent the limit for $\omega=0.80$ and $0.88$, respectively).  Dashed line indicates $\eta=0.0364$ and the covered light gray zone indicates the range suggested in (\ref{eta_s}).}
\end{figure}

As the last biased estimation task, we investigate the $\gamma$ estimation.  First we like to comment that the combination $f_{\chi}/f_{\beta}:=f_{\gamma}$ gives $-\gamma$ in the $M\to 0$ limit and, if $f_{\gamma}$ would allow us effective $\delta$-expansion, it would be a best function for our purpose.  However, $f_{\gamma}$ has a pole at $x=0.2388637\cdots$, which is confirmed by its diagonal Pad\'e approximants to high accuracy and this spoils the power of the $\delta$-expansion.  Thus, we use the estimated $\nu$ and $\eta$ to obtain $\gamma$ through $\gamma=2\nu(1-\eta/2)$.   In this protocol, two options are possible:  One is to substitute $\eta_{1}$ and the other is to substitute $\eta_{2}$.  Thus, we have two candidates of estimates, $\gamma_{1}=2\nu_{2}(1-\eta_{1}/2)$ and $\gamma_{2}=2\nu_{2}(1-\eta_{2}/2)$.  The result of estimates are summarized in Table 6 and Fig. 18.  Though the accurate estimate is not achieved yet, the trend of convergence is observed and the results at higher order implies agreement with the reference value (\ref{gamma_s}).   
\begin{table*}
\caption{$\omega$-biased estimation of $\gamma$ with the second order LDE.  In evaluation of $\gamma$ at order $N$ through $\gamma=2\nu(1-\eta/2)$, we used $\nu=\nu_{2}$ estimated in  the previous analysis at order $N+1$.  $\gamma_{1}=2\nu_{2}(1-\eta_{1}/2)$ and $\gamma_{2}=2\nu_{2}(1-\eta_{2}/2)$.}
\begin{center}
\begin{tabular}{ccccc}
\hline\noalign{\smallskip}
$order$ & 22 &  23 &  24 & 25    \\
\noalign{\smallskip}\hline\noalign{\smallskip}
$\gamma_{1}$ & 1.23833(108) & 1.23602(94) & 1.23732(87)  & 1.23686(100)  \\
$\gamma_{2}$ & 1.23579(91) & 1.23496(76) & 1.23676(86)  & 1.23646(64)  \\
\noalign{\smallskip}\hline
\end{tabular}
\end{center}
\end{table*}
\begin{figure}
\centering
\includegraphics[scale=0.8]{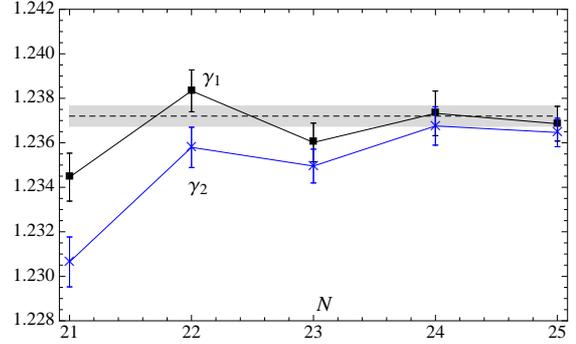}
\caption{(Color online) Plots of $\omega$-biased estimates of $\gamma$ via two protocols:  Black plot indicates the results via PMS conditions $\bar\Phi_{\chi }^{(1)}=0$ and $\bar\Phi_{\chi }^{(2)}\sim 0$ without using estimated $\nu$.  Blue plot indicates the result under the simple substitution $q_{2}=1/(2\nu_{2})$ with PMS condition on $\bar\Phi_{\chi>}$.  In both protocols, we used Fisher relation $\gamma=2\nu(1-\eta/2)$ to compute $\gamma$ with $\nu_{2}$.  Error bar denotes the variation range of estimated $\gamma$ under the change of $\omega$ from $0.80$ to $0.88$ (Lower and upper ends represents the limit for $\omega=0.80$ and $0.88$, respectively).  Dashed line indicates $\gamma=1.2372$ and the covered light gray zone indicates the range suggested in (\ref{gamma_s}).}
\end{figure}

To summarize this subsection, we find rough agreement with the reference results (\ref{eta_s}) and (\ref{gamma_s}).  In the estimation of both exponents, it seems that the PMS protocol is slightly superior to the simple substitution of pre-estimated $\nu$.  When $q_{2}$ does not obey in the included function some constraint, such as the self consistent one, it maybe  more suitable to be dealt with as an adjustable parameter, . 

\subsection{Unbiased estimation}
Here we attempt to estimate critical exponents in unbiased and self-contained manner.  We first address to the estimation of $\omega$ and $\nu$.  As many terms of the corrections are included in $f_{\beta(\chi)<}$, the estimation becomes accurate.  However, to fix associated unknown exponents, derivatives of target functions should be involved as many and then large order series in $1/M$ expansion is required.  Confined with $25$th order, it is safe to rely upon the two parameter ansatze, 
\begin{eqnarray}
f_{\beta<}&=&-\frac{1}{2\nu}+const\cdot x^{-q_{1}}+const\cdot x^{-q_{2}},\\
f_{\chi<}&=&\frac{\gamma}{2\nu}+const\cdot x^{-q_{1}}+const\cdot x^{-q_{2}}.
\end{eqnarray}

As the first attempt, let us discuss the extended PMS using conditions $\bar\Phi_{\chi> }^{(1)}=\bar\Phi_{\chi >}^{(2)}=0$ which gives $Q_{1\chi}(t)$ and $Q_{2\chi}(t)$ for the corresponding exponents $q_{1}$ and $q_{2}$.  Then, substituting them into the last condition that $\bar\Phi_{\beta>}$ becomes least sensitive at $t=t_{PMS}$ in the appropriate region, we obtain estimates of $q_{i}$ by $Q_{i\chi}(t_{PMS})=q_{i}^{*}$ $(i=1,2)$.  Since $q_{1}=\omega/2$ and $q_{2}=1/(2\nu)$, $Q_{i\chi}(t_{PMS})$ may give estimates of $\omega$ and $\nu$.  However, from an experience in $\beta_{c}$ estimation,  we have understood that $\nu_{PMS}:=\bar\Phi_{\beta >}|_{t=t_{PMS}, q_{i}=Q_{i\chi}(t_{PMS})}$ is more reliable for $\nu$ estimate (see (\ref{beta_lde1})).   The result reads, for example at $N=24$, $\nu_{PMS}=0.63537$, $\nu^{*}(=1/2q_{2}^{*})=0.41331$ and $\omega^{*}(=2q_{1}^{*})=1.37326$.  The results at other orders are also of the same level or worse.  The problem is found in the large discrepancy between the two estimates of $\nu$ by $\nu_{PMS}$ and $\nu^{*}$ (former is better than the later) and this would have some connection to largeness of estimated $\omega$.  The circumstance is quite similar when $\bar\Phi_{\beta>}^{(1)}=\bar\Phi_{\beta >}^{(2)}=0$ is used instead of $\bar\Phi_{\chi>}^{(1)}=\bar\Phi_{\chi >}^{(2)}=0$.  The pure PMS protocol is therefore not acceptable.  

As explicitly confirmed in the previous biased computation, one of the satisfactory features to be possessed in good protocols is an approximate realization of the self-consistency.  From this point of view, we consider that the best way is to employ the second protocol  (ii) presented in the former  subsection, which protocol is a hybrid of using PMS and the self-consistency in the combination of $\bar f_{\chi}$ and $\bar f_{\beta}$.  

Let us summarize here the conditions of the protocol (ii)-{\it unbiased} as follows:
\begin{itemize}
\item The self-consistency for the parameter $q_{2}$ is rigorously respected at the estimation point.
\item Use the stationarity condition $\bar\Phi_{\chi >}^{(1)}=0$ which gives $q_{2}\to Q_{2\chi}(t)$ as its effective function of estimation point $t$.
\item The upper limit of the observed scaling behavior of $Q_{2\chi}(t)$ is matched with the self-consistent point.
\end{itemize}

According to the above protocol, the estimation goes as follows:  For example, consider the case at $N=24$, the highest order of $\Phi_{\beta}$.  Taking three conditions into account, we impose the coupled equations
\begin{eqnarray}
\bar\Phi_{\beta>}|_{q_{2}=Q_{2\chi}}&=&-Q_{2\chi},\\
Q_{2\chi}^{(1)}&=&0.
\label{stationary-q2}
\end{eqnarray}
The second equation comes from the interpretation that the upper limit of the scaling of $Q_{2\chi}$ would be attained at the stationary point.  
We then obtain the solutions, \\$t=0.11793799038\cdots=t_{SC}$ and $\omega=0.7451128251\cdots=\omega_{SC}$.  Then, the substitution of the solution into $\bar\Phi_{\beta>}|_{Q_{2\chi}}$ or $Q_{2\chi}$ yields $\nu=1/(2q_{2})=0.628830753\cdots=\nu_{SC}$.  Next, consider the case at $N=23$.  At this order the set of solutions is absent and we loosen the second condition such that $Q_{2\chi}$ should approximately satisfy the stationarity at the self-consistent point $t=t_{SC}$; That is, we instead use $Q_{2\chi}^{(2)}=0$ at $t=t_{SC}$ (Above $20$th orders, we found that at even orders $t_{SC}$ is obtained as the stationary point of $Q_{2\chi}$ and at odd orders as the least variation point in the scaling region.).   This modification then gives $t_{SC}=0.11673398933\cdots$ and $\omega_{SC}=0.7120047317\cdots$.  Thus, we obtain $\nu_{SC}=0.6278846723\cdots$.  For the graphical representation of the situation, see Fig. 19 and Fig. 20.   
\begin{figure}
\centering
\includegraphics[scale=0.8]{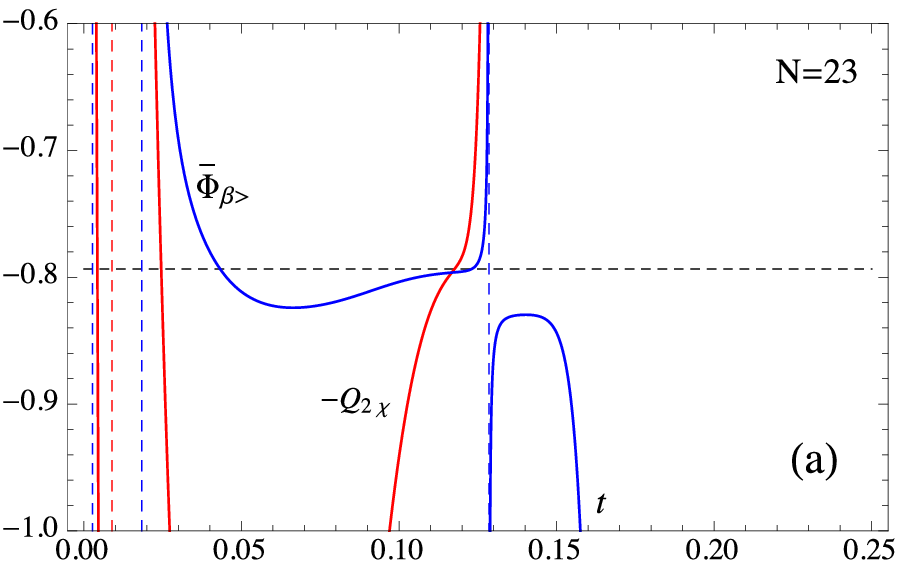}
\includegraphics[scale=0.8]{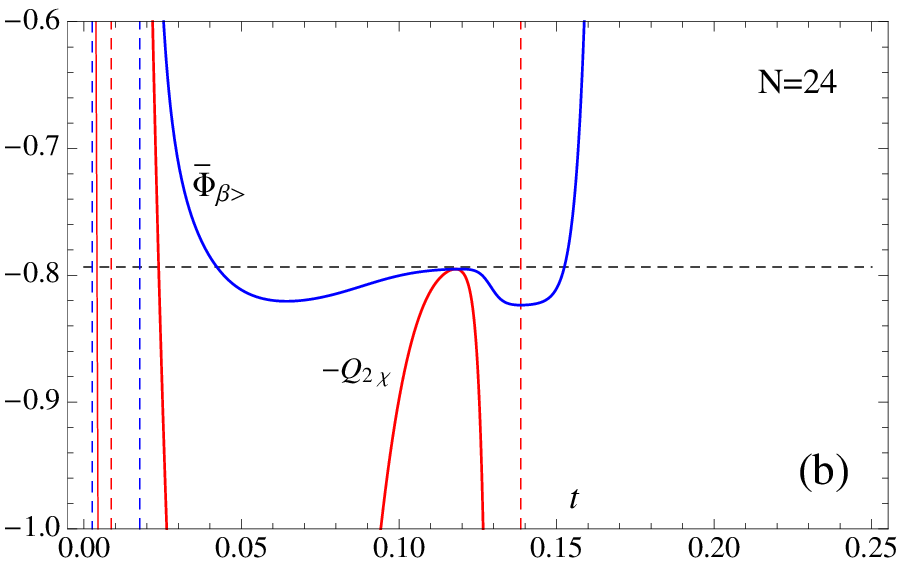}
\caption{(Color online) $\bar\Phi_{\beta >}|_{Q_{2\chi}(t)}$ and $-Q_{2\chi}(t)$ as functions of $t$ which specifies the solution of stationarity condition $\Phi_{\chi<}^{(1)}=0$.  Compare with plots in FIG. 13(c),(d).   Horizontal dashed lines indicate the standard value of $-1/(2\nu)=-1/(2\times 0.6301)$.  The common point of the two curves indicates the acceptable self-consistent solution.}
\end{figure}
\begin{figure}
\centering
\includegraphics[scale=0.8]{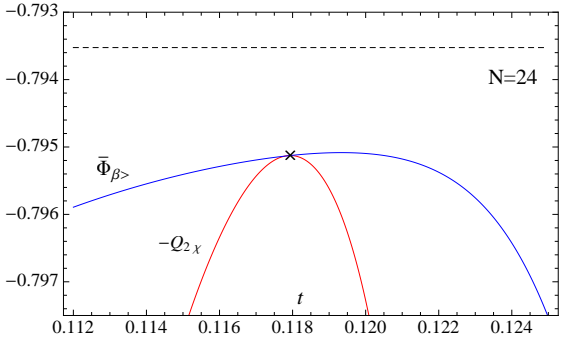}
\caption{(Color online) $\bar\Phi_{\beta>}|_{Q_{2\chi}(t)}$ and $-Q_{2\chi}(t)$ in the vicinity of the self-consistent solution represented by the cross mark at $24$th order.  Horizontal dashed line indicates the standard value of $-1/(2\nu)=-1/(2\times 0.6301)$.}
\end{figure}

The results from $20$th to $24$th order are summarized in Table 7 and Fig. 21.   We observe that both of $\omega_{SC}$ and $\nu_{SC}$ are monotonically increasing with the order $N$.  For $\omega$ results, this may explain the smallness of $\omega_{SC}$ such that the incorporation of un-computed higher order terms would lift up the higher order estimates near (\ref{omega_s}).   The estimate of $\nu$ at the present highest order is slightly smaller than (\ref{nu_s}) but the sequence $\{\nu_{SC}\}$ indicates the value $\sim 0.63$ and essentially consistent in the world average.  
\begin{table}
\caption{Unbiased estimation of $\omega$ and $\nu$ via the self-consistent PMS protocol (ii)-{\it unbiased}.}
\begin{center}
\begin{tabular}{cccccc}
\hline\noalign{\smallskip}
$order$ & 20 &  21 & 22 &  23 & 24    \\
\noalign{\smallskip}\hline\noalign{\smallskip}
$\omega_{SC}$ & 0.31643 & 0.41711 & 0.61357  & 0.71200 & 0.74513  \\
$\nu_{SC}$ & 0.60424 & 0.61507 & 0.62526  & 0.62788 & 0.62883  \\
\noalign{\smallskip}\hline
\end{tabular}
\end{center}
\end{table}

\begin{figure}
\centering
\includegraphics[scale=0.8]{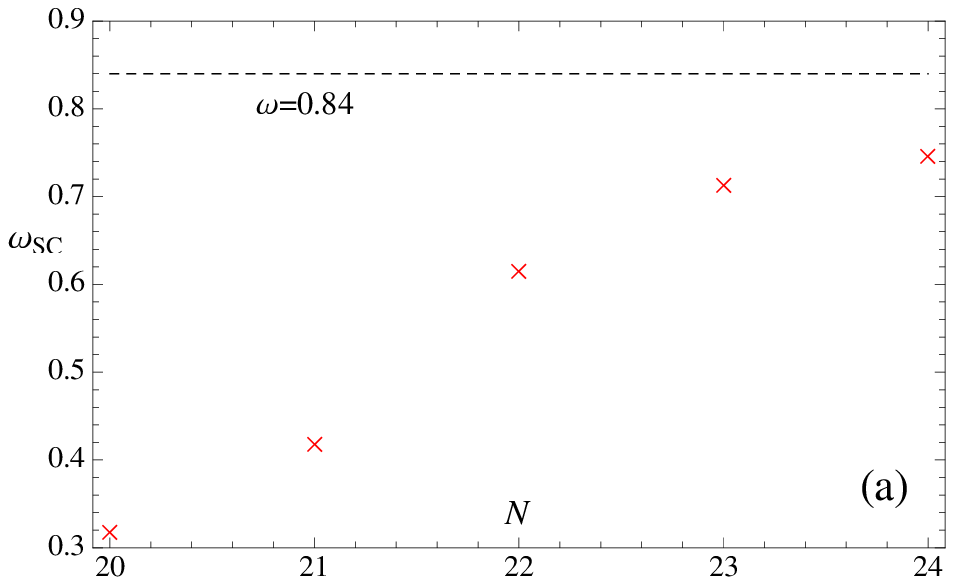}
\includegraphics[scale=0.8]{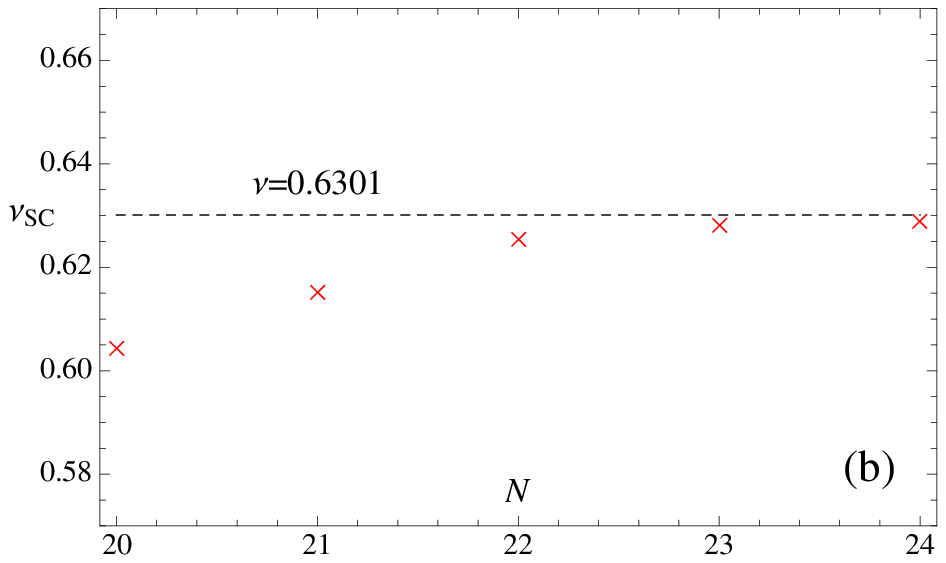}
\caption{(Color online) Plots of estimated $\omega$ (plot (a)) and $\nu$ (plot (b)) via the protocol (ii)-{\it unbiased}.}
\end{figure}

Next, we turn to estimate $\eta$ and then $\gamma$.  We use $\omega_{SC}$ just obtained in the protocol (ii)-{\it unbiased}.  The estimate of $\nu$ is not required since it is desirable to avoid as possible the bias from the previous estimation.  Then, we use (\ref{q2chi}) to substitute it into the second derivative of $\bar\Phi_{\chi >}$ and select the point satisfying
\begin{equation}
\bar\Phi_{\chi >}^{(2)}\sim 0
\end{equation}
where $\sim$ means the exact zero (at odd orders) or approximate zero (even orders).  Having obtained the solution $t_{PMS}$, we simply substitute it into $\bar \Phi_{\chi>}$ and identify
\begin{equation}
\frac{\gamma}{2\nu}=1-\frac{\eta}{2}=\bar \phi_{\chi>}\Big |_{q_{2}=Q_{2\chi}(t_{PMS}), t=t_{PMS}}.
\end{equation}  
We obtained the results shown in Table 8 and Fig. 22 (a).  The estimate of $\gamma$ is simply done through $\gamma=2\nu(1-\eta/2)$.  See the result in Table 8 and Fig. 22 (b).  We first remark that, at $N=25$, $t_{PMS}$ exactly agrees with $t_{SC}$ obtained in the $\nu$ estimation.   This is because the second stationarity condition $\bar\Phi_{\chi>}^{(2)}=0$ is equivalent with $Q_{2\chi}^{(1)}=0$ (see (\ref{stationary-q2})).  When $\bar\Phi_{\chi>}^{(2)}\neq 0$ at $t_{PMS}$, $t_{PMS}$ shows just a slight difference from $t_{SC}$.  These facts mean that the value of $Q_{2\chi}$ at $t_{PMS}$ yields exactly the same or very close value to $\nu_{SC}$ at respective orders (Remind that $Q_{2\chi}$ converges to $1/(2\nu)$ in the scaling limit).   The protocol (ii)-{\it unbiased} thus fulfills the unified treatise of estimations of various exponents. 
\begin{table}
\caption{Estimation of $\eta$ and $\gamma$ biased by $\omega_{SC}$ obtained in the self-consistent PMS protocol (ii)-{\it unbiased}.  Lists of two kinds of $\gamma$ come from the following :  Upper one denoted by $\gamma_{SC}$ shows the result obtained by use of the estimated $\nu_{SC}$ and the lower one represented by the asterisk by use of $\nu=0.6301$ quoted in (\ref{nu_s}) from ref. \cite{peli}.  The later sequence is added to guide how estimate of $\gamma$ is affected by $\nu$-value.}
\begin{center}
\begin{tabular}{cccccc}
\hline\noalign{\smallskip}
$order$ & 21 &  22 & 23 &  24 & 25    \\
\noalign{\smallskip}\hline\noalign{\smallskip}
$\eta_{SC}$ & 0.01186 & 0.02474 & 0.03355  & 0.03578 & 0.03633  \\
$\gamma_{SC}$ & 1.20131 & 1.21869 & 1.22954  & 1.23387 & 1.23481  \\
$\gamma^{*}$ & 1.25272 & 1.24461 & 1.23906  & 1.23766 & 1.23731  \\
\noalign{\smallskip}\hline
\end{tabular}
\end{center}
\end{table}
\begin{figure}
\centering
\includegraphics[scale=0.8]{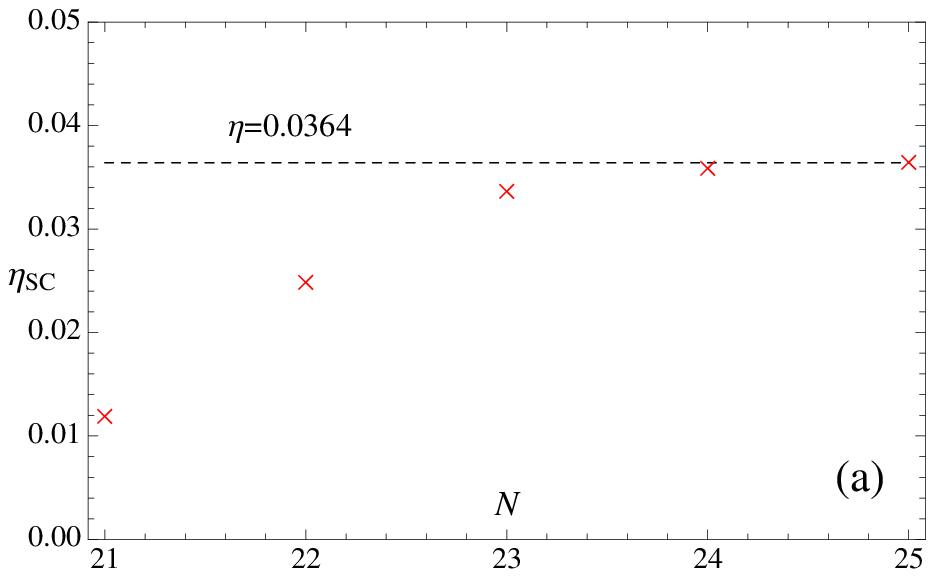}
\includegraphics[scale=0.8]{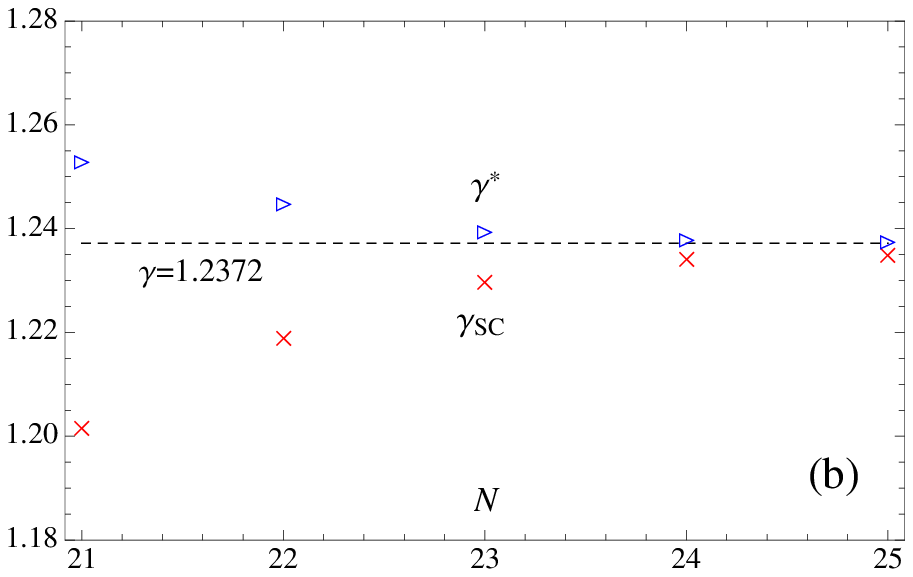}
\caption{(Color online) Plots of estimated $\eta$ (plot (a)) and $\gamma$ (plot (b)) biased by $\omega_{SC}$ obtained in the self-consistent PMS protocol (ii)-{\it unbiased} applied to $f_{\chi>}$.  Red cross represents the data $\gamma_{SC}$ and blue triangle the data $\gamma^{*}$.}
\end{figure}

The estimate of $\eta$ at $25$th order agrees excellently with the quoted value  (\ref{eta_s}) \cite{peli}.  The lists of the $\gamma$ estimation are composed by two sequences.  The upper sequence stands for the results by use of $\nu_{SC}$ shown in Table 7.  The lower sequence asterisked shows the results by use of $\nu$ quoted in (\ref{nu_s}) as the standard \cite{peli}.  The result of $\gamma$ at $N=25$ is slightly lower than the standard one (\ref{gamma_s}) though the trend of the sequence implies acceptable limit.  The list of $\gamma$ in the upper sequence is biased from $\nu_{SC}$ and it would be interesting to see that the use of (\ref{nu_s}) gives excellent agreement with the standard $\gamma$ quoted in (\ref{gamma_s}).  This indicates that our approach to the critical exponent estimation is totally satisfactory.

\section{Concluding remarks}
To summarize the paper, we have investigated the estimation of critical quantities in expansion in a mass argument.  First, the critical inverse temperature $\beta_{c}$ has estimated through LDE with constant coefficients related to the critical exponents.   With the aid of the $\delta$-expansion and the extended PMS condition, we have obtained at the highest order $25$th, $\beta_{c}=0.221687$, larger than the standard upper limit $\sim 0.22166$.  As a by product, we have also obtained the rough estimate of $\nu$ at third order LDE.  

Independently, the improved estimations of $\nu$ and other critical exponents have been attempted by employing $\beta^{(2)}/\beta^{(1)}$ and $\chi^{(1)}/\chi$ as the target functions.  Under the hybrid protocol of PMS and self-consistency point of view, we have attempted biased and un-biased estimations of critical exponents.  Biased by $\omega=0.84(4)$, we obtained results in good agreement of world averages.  In unbiased approach, satisfactory results are obtained for $\nu$, $\eta$ and $\gamma$ in self-contained way;  The best estimate of $\omega$ is $0.74513$ and smaller than the range indicated in ref. \cite{peli}, though the obtained sequence $\{\omega_{SC}\}$ may probably grow to the value around $\sim 0.8$ at large enough orders.   Exponent $\nu$ has estimated to be $0.62883$, slightly smaller compared to (\ref{nu_s}).  In contrast, the estimate $\eta=0.03633$ agrees very well with (\ref{eta_s}).  We emphasize that in our approach, the scaling behavior is directly observable by the numerical plots of relevant functions.  This feature enables us to identify which solution among the ones produced by the extended PMS or self-consistency condition should be really relied upon.  

The works of Guttmann, Butera and Comi \cite{gut,butera1}, using solely the high temperature expansion with the aid of differential approximants method, used shorter series and the direct comparison is not possible.  We just mention that our estimates of $\nu$ and $\gamma$ are better than their results.  In \cite{gut}, $\gamma$ is estimated as $1.2431(24)$ and in \cite{butera1}, $\nu=0.634(2)$ and $\gamma=1.244(3)$.  However, their estimates of $\beta_{c}$ is better than ours; $\beta_{c}=0.221657(7)$ in \cite{gut} and $0.221663(9)$ in \cite{butera1}.  In fact, we have expected that the improved estimate of critical exponents may help to improve the $\beta_{c}$ estimation.  However, we have not obtained yet a solid scheme for the essential improvement.  

Finally, we remind that the estimation protocol (ii)-{\it unbiased} fulfills the conditions of self-consistency in $\bar\Phi_{\beta}$ exactly and the approximate (or exact in cases) stationarity of $Q_{2\chi}$ and $\bar\Phi_{\beta}$ at the estimation point.  These are also valid for $\bar\Phi_{\chi}$ function for $\eta$ estimation.  We note that those conditions should be respected when the functions under the investigation are really in the scaling region.    
The set of those conditions is strong and tightly bounds the estimation procedures.  If one can find more flexible variant, the accuracy of estimates might be improved.  



\begin{thebibliography}{99}
\bibitem{comment}  For example, when the site $j$ is on the diagonal direction with $j=J(1,1,1)$, it is found at the leading order that $<s_{0}s_{j}>\sim \frac{(3J)!}{(J!)^3}(\tanh \beta)^{3J}$.  At large enough $J$, $\lim_{J\to \infty}(\frac{(3J)!}{(J!)^3})^{1/3J}=3$ and $<s_{0}s_{j}>\sim \exp[3J\log 3 \beta]$.  Thus we obtain $\xi\sim \sqrt{3}\log (3M)$.
\end{thebibliography}
\end{document}